\documentclass[pre,aps,10pt,superscriptaddress,twocolumn,floatfix]{revtex4-1}

\usepackage[nice]{nicefrac}
\usepackage{graphicx,color}
\usepackage{amsmath,amssymb,bm,bbm}
\usepackage{amsmath}
\usepackage{braket}

\usepackage[plainpages=false,pdfpagelabels,colorlinks=true,linkcolor=red,urlcolor=blue,citecolor=blue,pdftitle={},pdfauthor={},pdfdisplaydoctitle=true,pdfduplex=DuplexFlipLongEdge]{hyperref}

\definecolor{darkred}{rgb}{0.90,0.2,0.2}
\definecolor{darkgreen}{rgb}{0,0.60,.2}
\definecolor{darkblue}{rgb}{0.1,0.3,1}
\definecolor{grey}{cmyk}{0,0,0,0.25}
\definecolor{orange}{cmyk}{0,0.6,0.8,0}

\begin{document}

\title{Scale-invariant critical dynamics at eigenstate transitions}

\author{Miroslav Hopjan}
\affiliation{Department of Theoretical Physics, J. Stefan Institute, SI-1000 Ljubljana, Slovenia}
\author{Lev Vidmar}
\affiliation{Department of Theoretical Physics, J. Stefan Institute, SI-1000 Ljubljana, Slovenia}
\affiliation{Department of Physics, Faculty of Mathematics and Physics, University of Ljubljana, SI-1000 Ljubljana, Slovenia\looseness=-1}

\begin{abstract}
The notion of scale-invariant dynamics is well established at late times in quantum chaotic systems, as illustrated by the emergence of a ramp in the spectral form factor (SFF).
Building on the results of the preceding Letter [\href{https://doi.org/10.1103/PhysRevLett.131.060404}{Phys.~Rev.~Lett.~\textbf{131}, 060404 (2023)}], we explore features of scale-invariant dynamics of survival probability and SFF at criticality, i.e., at eigenstate transitions from quantum chaos to localization.
We show that, in contrast to the quantum chaotic regime, the quantum dynamics at criticality do not only exhibit scale invariance at late times, but also at much shorter times that we refer to as {\it mid-time dynamics}.
Our results apply to both quadratic and interacting models.
Specifically, we study Anderson models in dimensions three to five and power-law random banded matrices for the former, and the quantum sun model and the ultrametric model for the latter, as well as the Rosenzweig-Porter model.
\end{abstract}
\maketitle

\section{Introduction}

Since the pioneering paper by M. Berry~\cite{berry_85}, the spectral form factor (SFF) is considered as one of the key diagnostics of quantum chaos.
Specifically, one says that the dynamics of a quantum system is chaotic if the SFF after sufficiently long time follows the so-called ramp, i.e., a linear increase in time that is predicted by the random matrix theory (RMT)~\cite{mehta_91}.
The SFF is commonly normalized such that different quantum chaotic systems follow the same ramp, irrespective of the system size~\cite{suntajs_bonca_20a, suntajs_prosen_21, suntajs_vidmar_22}.
Emergence of such a scale-invariant dynamical behavior represents a convenient tool to detect universality of chaotic dynamics, as shown by a great amount of recent studies~\cite{Kos18, Chan18, Friedman19a, suntajs_bonca_20a, suntajs_prosen_21, suntajs_vidmar_22, Liu17, Gharibyan18, Vasilyev20, Winer20, Liao20, Prakash21, Moudgalya21, Dieplinger21, Winer22, Winer22b, Joshi22, Winer22a, Winer23a, loizeau23, Dag23, Barney23, Fritzsch23a}.
The concept of SFF can be extended to survival probabilities~\cite{Peres84} of initial nonequilibrium wavefunctions, which, at least within the RMT, also exhibit scale-invariant dynamics at long times~\cite{mehta_91, Santos2017, torresherrera_garciagarcia_18, Schiulaz_19}.
Of particular importance is the time of the onset of scale-invariant chaotic behavior in the SFF, denoted as the Thouless time $t_{\rm Th}$.
However, $t_{\rm Th}$ may be very long and it typically increases with system size.
For example, in systems of linear size $L$ that exhibit diffusion, one expects $t_{\rm Th} \propto L^2$~\cite{Thouless74, Gharibyan18, Friedman19a, suntajs_bonca_20a, Sierant2020, Colmenarez22}.

The main focus of this paper are the dynamical transitions from chaos to localization, for which critical behavior in finite systems is associated with an abrupt transition in eigenstate properties.
The eigenstate transitions are relatively well understood in systems described by quadratic Hamiltonians, with Anderson localization being a paradigmatic example~\cite{anderson_58, Abrahams79, Aubry80, Suslov82, EversMirlin2008}.
Recently, a new class of interacting models based on the avalanche theory~\cite{deroeck_huveneers_17} was introduced, with the potential to provide a stepping stone into a more detailed understanding of critical behavior at ergodicity breaking phase transitions~\cite{deroeck_huveneers_17, deroeck_imbrie_17, luitz_huveneers_17, thiery_huveneers_18, goihl_eisert_19, gopalakrishnan_huse_19, potirniche_banerjee_19, crowley_chandran_20, sels_22, morningstar_colmenarez_22, suntajs_vidmar_22, crowley_chandran_22b, hopjan2023, pawlik_sierant_23}.

Motivated by the emergence of scale-invariant dynamics in quantum chaotic systems, we here ask two questions.
First, does some notion of scale-invariant dynamics persist at criticality?
If the answer is affirmative, in which time regimes may one then expect scale-invariant behavior?
In particular, does one need to wait for very long times, e.g., to the Thouless time, to observe scale invariance, or does it emerge already at much shorter times?

\begin{figure}[!t]
\centering
\includegraphics[width=0.98\columnwidth]{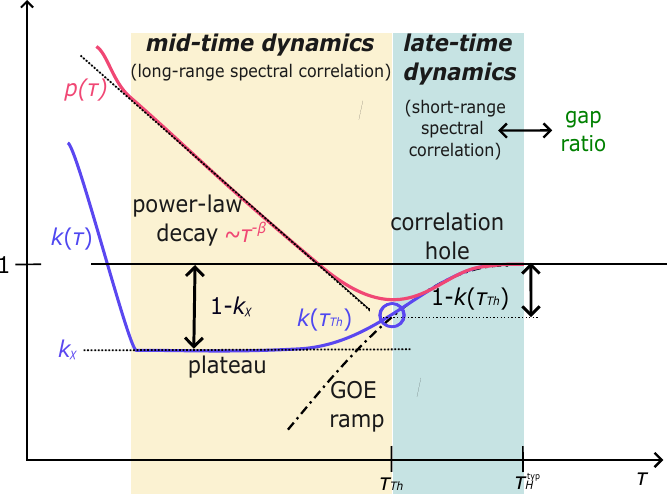}
\caption{Concept of mid-time and late-time dynamics at eigenstate transition, for the dynamics of the SFF $k$ and survival probability $p$ of initially site-localized states. 
We measure time $t$ in units of typical Heisenberg time $t_{H}^{\rm typ}$, such that $\tau=t/t_{H}^{\rm typ}$, and we normalize $k$ and $p$ such that they approach $k=p=1$ at $\tau>1$.
The late-time dynamics is defined as the dynamics in between the scaled Thouless time $\tau_{\rm Th}=t_{\rm Th}/t_{H}^{\rm typ}$ and the scaled typical Heisenberg time $\tau_{H}^{\rm typ}=1$. 
At the transition, the late-time dynamics is preceded by the scale-invariant mid-time dynamics, where $k$ exhibits a broad plateau, $k\approx k_\chi$, and $p$ decays with power-law dependence $p\approx a_0\, \tau^{-\beta} $.
While the late-time dynamics are universal in quantum chaotic sense, the properties of mid-time dynamics, such as the values of $K_\chi$ and $\beta$, depend on the properties of transition point. 
}
\label{fig_1}
\end{figure}

The main result of this paper and the preceding Letter~\cite{hopjan2023} is that for both quadratic systems exhibiting Anderson localization, and interacting systems exhibiting ergodicity breaking phase transition, the answer to the first question is indeed affirmative, and that scale invariance at criticality emerges already in mid-time dynamics, as sketched in Fig.~\ref{fig_1}.
The central measures of scale invariance are the SFF $k$, which exhibits a broad plateau in the mid-time dynamics, and the survival probability $p$ of site-localized states, which exhibits a power-law decay.
Importantly, both quantities are measured in units of the typical Heisenberg time (proportional to the typical inverse level spacing), and are normalized such that they also exhibit the late-time scale invariance in the chaotic regime.
They hence represent useful indicators of both chaotic dynamics deep in the chaotic regime as well as the critical dynamics at the boundary of quantum chaos.

We note that the main results of this paper and the preceding Letter~\cite{hopjan2023} is to establish the link to the critical dynamics of interacting systems at the ergodicity breaking transition point.
In the context of quadratic systems, many studies contributed to the identification of certain properties that are responsible for the emergence of critical dynamics.
These studies range from the observation of scale invariance in eigenstate correlations$~$\cite{Chalker88}
and in short-range spectral statistics$~$\cite{Shklovskii93}, to studies of long-range spectral statistics, in particular their spectral rigidity~\cite{Kravtsov94,Aronov95,Mirlin00,Evers2000}. The long-range spectral correlations were further argued to be related to the eigenfunction correlations at criticality in Refs.$~$\cite{Chalker96a,Chalker96b,Mirlin00,Evers2000}, where the mid-time features of the SFF and survival probability were also discussed$~$\cite{Chalker96b}. The later insights were sharpened by noting that the emergence of a broad scale-invariant plateau in the SFF of the Anderson models is a strong indicator of criticality~\cite{suntajs_prosen_21, suntajs_prosen_23}.
An important connection to interacting models was recently made by observing a very similar plateau in the SFF of the interacting quantum sun (QS) model at the ergodicity breaking transition point~\cite{suntajs_vidmar_22}.

Here we follow the perspective that the SFF is a survival probability of a special initial state$~$\cite{delcampo_molinavilaplana_17}, which is an equal superposition of Hamiltonian eigenstates.
It is then natural to seek for scale invariance also in survival probabilities of other initial states.
In quadratic systems, several studies contributed to the understanding that the survival probability of the initial site-localized states exhibits a power-law decay in the vicinity of Anderson transition$~$\cite{Ketzmerick_92,Brandes96,Ohtsuki97}.
This observation paved way towards scale invariance in mid-time dynamics, as well as towards making the connection of the power-law exponent with the wavefunction fractal dimension$~$\cite{Huckestein94,Ketzmerick_97,Ohtsuki97}.
The first contribution of our Letter~\cite{hopjan2023} was to introduce the scaled survival probability $p$ mentioned above, which exhibits scale invariance in both mid- and late-time dynamics at criticality.
The later quantity allows for establishing the connection between the power-law exponent and the wavefunction fractal dimension also in other quadratic models such as the Aubry-André model.
The main contribution of~\cite{hopjan2023} was then to generalize these findings to the QS model of the avalanche theory, which exhibits scale-invariant mid-time and late-time dynamics at the ergodicity breaking transition point.
In this paper we first provide a comprehensive overview of the subject, and then go beyond the Letter~\cite{hopjan2023} by studying a broad range of different models, ranging from the quadratic models such as the Anderson models in dimensions three to five and power-law random banded matrices, to the interacting models such as the QS model and the ultrametric model defined within the avalanche picture, as well as the Rosenzweig-Porter model {that we consider as a separate class of models}.
Moreover, we also explore to which degree scale invariance in survival probabilities emerges for other initial states, such as plane waves, and we find that it indeed exhibits certain fingerprints of scale invariance, in particular at weak multifractality.

The paper is organized as follows.
In Sec.~\ref{sec:models} we define the models studied in this paper.
In Sec.~\ref{sec:methods} we introduce the measures of quantum dynamics, i.e., the SFF and survival probabilities, and we comment on their similarities.
In Sec.~\ref{sec:scaleinvariant} we discuss the key properties of scale-invariant dynamics in the chaotic regime and at criticality, and we introduce the scaled survival probabilities that are the central object of investigation.
Numerical results for the quadratic models, interacting models and the Rosenzweig-Porter model are presented in Secs.~\ref{sec:Anderson_PLRB},~\ref{sec:Avalanche} and~\ref{sec:RPmodel}, respectively.
In particular, we establish similarity between the quadratic and interacting models, and we highlight subtle differences present in the Rosenzweig-Porter model.
We summarize our findings in Sec.~\ref{sec:mid_vs_late}, and we conclude in Sec.~\ref{sec:conclusions}.

\section{Models} \label{sec:models}

In this section we introduce the models that exhibit eigenstate transitions and whose critical dynamics are studied in this paper. 
We split them into two main categories, quadratic models and interacting models, and finally we also consider the Rosenzweig-Porter model as a separate category of models.

\subsection{Quadratic models}

We start with the well-known Anderson model$~$\cite{anderson_58} on a $d$-dimensional hypercubic lattice of linear size $L$,
\begin{equation}
\label{eq:ham3DA}
\hat H= -{J}\sum_{\langle ij\rangle}^{} (\hat{c}_{i}^{\dagger}\hat{c}_{j}^{}+ \hat c_j^\dagger \hat c_i) + \sum_{i=1}^{D}\epsilon_{i}\hat{n}_{i}^{}\;,
\end{equation}
where $\hat{c}_{j}^{\dagger}$ ($\hat{c}_{j}^{}$) are the fermionic creation (annihilation) operators at site $j$, $J$ is the hopping matrix element between nearest neighbor sites,
$\hat{n}_{i}^{}=\hat{c}_{i}^{\dagger}\hat{c}_{i}^{}$ is the site occupation operator, and $\epsilon_{i}$ are the on-site energies.
The later are independent and identically distributed random variables, drawn from a box distribution $\epsilon_i \in [-W/2,W/2]$. {We consider periodic boundary conditions.}
Since the model is quadratic, the number of lattice sites coincides with the single-particle Hilbert space dimension, $D=L^d$. 

Theoretical arguments suggest the transition in this model to occur in dimensions $d>2$~\cite{Abrahams79}. In three dimensions (3D), numerical studies of transport properties of single-particle eigenstates in the center of the energy band$~$\cite{kramer_mackinnon_93, MacKinnon81, MacKinnon83}, based on the transfer-matrix technique, showed that the system is insulating at $W >W_{c} \approx 16.54 J$$~$\cite{Ohtsuki18}, and at $W<W_{c}$ it becomes diffusive~\cite{Ohtsuki97, Zhao20, Herbrych21}.
The value of the transition point $W_{c}$ grows with the dimension $d$ and was accurately calculated using numerical techniques~\cite{Garcia-Garcia07, Rodriguez10, Pietracaprina16, Mard17, Tarquini17, Herbrych21, Sierant2020, suntajs_prosen_21}.
At the transition, the model exhibits subdiffusion~\cite{Ohtsuki97} and multifractal single-particle  eigenfunctions$~$\cite{EversMirlin2008, Rodriguez09, Rodriguez10}. The transition point is energy dependent, i.e., at $W >W_{c}$ all single-particle states are localized in site-occupation basis, while at $W <W_{c}$ the system exhibits a mobility edge$~$\cite{Schubert_05}.

Next, we introduce the ensemble of power-law random banded (PLRB) matrices.
We define the corresponding PLRB Hamiltonian as a quadratic Hamiltonian, whose matrix elements in the single-particle Hilbert space of size $D=L$ are given by
\begin{equation}
\label{eq:ham_ensemble}
\hat H= \sum_{i,j=1}^{L}  h_{i,j} \hat{c}_{i}^{\dagger}\hat{c}_{j}^{},\;\;
h_{i,j}=h_{j,i}=\frac{\mu_{i,j}}{[1+(|i-j|/b)^{2a}]^{1/2}}\;,
\end{equation}
i.e., the matrix elements decrease as power laws with distance $|i-j|$~\cite{Mirlin96, Evers2000, Mirlin00, EversMirlin2008}. 
In Eq.~(\ref{eq:ham_ensemble}), $\mu_{i,j}$ are independent and identically distributed random variables drawn from a box distribution, $\mu_{i,j} \in  [-1,1]$~\cite{Bera18}, but other distributions can also be considered~\cite{Mirlin96, Evers2000, Mirlin00}.

The PLRB model is parameterized by two parameters $a$ and $b$, and it exhibits an eigenstate transition at $a=1$ for all $b$.
The single-particle eigenstates are delocalized at $a<1$ and localized in site-occupation basis at $a>1$.
The model was designed as a toy model to understand the features typical for the Anderson model close to and at criticality.
At the critical value $a=1$ the system exhibits multifractality$~$\cite{Levitov90,Mirlin96,Evers2000,Mirlin00} and spectral statistics are intermediate between the Wigner-Dyson and Poisson statistics~\cite{Mirlin96, Evers2000, Mirlin00}.
These properties are tuned by the parameter $b$, ranging from strong multifractality at $b\ll1$ to weak multifractality at $b\gg1$.
Some of the dynamical aspects of critical ensembles were considered both analytically$~$\cite{Kravtsov10,Kravtsov11,Kravtsov12} and numerically$~$\cite{Bera18}.

\subsection{Interacting models}

In the domain of interacting models we consider two representatives, which are both expected to describe the avalanche mechanism of ergodicity breaking phase transitions.

The first model, dubbed the quantum sun (QS) model ~\cite{suntajs_vidmar_22, hopjan2023, pawlik_sierant_23, suntajs2023similarity}, shares many similarities with the initially proposed toy model of quantum avalanches~\cite{deroeck_huveneers_17, luitz_huveneers_17}.
The model consists of $N+L$ spin-1/2 degrees of freedom in a Fock space of dimension $D=2^{N+L}$. It is divided into a quantum dot with $N$ spins and a remaining subsystem with $L$ spins outside the dot, described by the Hamiltonian
\begin{equation}
\label{eq:hamQSM}
\hat H= \hat R+g_0\sum_{i=0}^{L-1} \alpha^{u_i} \hat{S}_{n_i}^{x}\hat{S}_{i}^{x}+\sum_{i=0}^{L-1}h_{i}\hat{S}_{i}^{z} \;.
\end{equation}
The interactions within the dot, denoted by $\hat R$, are all-to-all and they exclusively act on the dot subspace.
They are represented by a $2^N \times 2^N$ random matrix drawn from the Gaussian orthogonal ensemble (GOE)~\footnote{Random matrix $R$ is drawn from the Gaussian orthogonal ensemble:
$R=\eta(A+A^T)/2 \in \mathbb{R}^{2^N\times 2^N}$, where $A_{ij}= {\rm norm}(0,1)$ are random numbers drawn from a normal distribution with zero mean and unit variance. We set $\eta=0.3$ as in Refs. \cite{luitz_huveneers_17, suntajs_vidmar_22, hopjan2023}. The resulting matrix is embedded to matrix of the full dimension $2^{N+L}$ by Kronecker product $\hat{R}=R \otimes \mathcal{I}$ where $\mathcal{I}$ is the identity matrix of dimension $2^L$.}.
The spins outside the dot are subject to local magnetic fields $h_i \in [0.5,1.5]$ that are drawn from a box distribution.
Each of the spins outside the dot is coupled to one spin in the dot, with the interaction strength $\alpha^{u_i}$.
For a chosen spin $i$ outside the dot, we randomly select an in-dot spin $n_i$.
The coupling to the first spin outside the dot ($i=0)$ is set to one since $u_0=0$, while at $i \geq 1$, $u_i \in [i-0.2, i+0.2]$ is drawn from a box distribution. 

The ergodicity breaking transition in the QS model is well established both analytically and numerically~\cite{deroeck_huveneers_17, luitz_huveneers_17, suntajs_vidmar_22, hopjan2023, pawlik_sierant_23, suntajs2023similarity}.
The analytical prediction for the transition point is $\bar\alpha=1/\sqrt{2}\approx 0.707$~\cite{deroeck_huveneers_17, suntajs2023similarity}, while numerical studies usually observe the transition at similar, but slightly larger values of critical points $\alpha=\alpha_c$~\cite{suntajs_vidmar_22, hopjan2023, pawlik_sierant_23, suntajs2023similarity}.
Here we follow Ref.~\cite{hopjan2023}, which set $N=5$ and $g_0=1$ in Eq.~(\ref{eq:hamQSM}) and estimated $\alpha_c\approx 0.716$ from the gap ratio statistics.
Eigenstates of the QS model in the middle of the spectrum exhibit ergodicity at $\alpha>\alpha_c$, the critical point $\alpha=\alpha_c$ is multifractal, and the nonergodic phase at $\alpha<\alpha_c$ exhibits Fock space localization~\cite{suntajs2023similarity}.

The second model is the ultrametric (UM) model.
This model was initially introduced in the single-particle picture to study Anderson localization on ultrametric lattices~\cite{fyodorov2009anderson, rushkin2011universal, vansoosten2018phase, Bogomolny2011, MendezBermudez2012, Gutkin2011, vonSoosten2017, Bogomolny2018, vonSoosten2019}.
However, recent study has highlighted similarities between the UM model and the QS model when defined for interacting systems in a Fock space of dimension $D$~\cite{suntajs2023similarity}.
Following the convention in the QS model, we define a Fock space of $N+L$ spin-1/2 particles, with $D = 2^{N+L}$.
The model Hamiltonian is constructed as a sum of block-diagonal random matrices $\hat{H}_k$ with $k=0, 1, \ldots, L.$ At each $k,$ the matrix structure of $\hat{H}_k$ consists of $2^{L - k}$ diagonal blocks of size $2^{N + k} \times 2^{N + k}.$ We sample each random block independently from the GOE distribution and use normalization such that 
\begin{equation}
    H^{(i)}_k = \frac{R^{(i)}}{\sqrt{2^{N + k} + 1}}, \hspace{5mm} i = 1, \ldots, 2^{L - k}\;,
\end{equation}
see~\cite{suntajs2023similarity} for details.
Here, the superscript $i$ denotes the $i$th random block. We sample its matrix elements in analogy to the QS model, hence $R^{(i)} = \frac{1}{\sqrt{2}}(A + A^T) \in 2^{N + k} \times 2^{N+k}.$ 
The full Hamiltonian of the UM model then reads
\begin{equation}\label{eq:def_rmt_model}
    \hat{H} = \hat{H}_0 + J_c\sum\limits_{k=1}^L \alpha^k \hat{H}_k, \hspace{5mm} \alpha\in [0, 1).
\end{equation}
The first term $\hat{H}_0$ of size $2^{N} \times 2^{N}$ describes the initial quantum dot that consists of $N$ spin-1/2 particles in the absence of coupling to the external particles. The sum in the second term mimics the exponentially decaying coupling between the dot and the $k$-th localized spin-1/2 particle through the exponentially decaying values of $\alpha^k.$ Additionally, we have also included the parameter $J_c$ tuning the overall perturbation strength, which carries certain analogies with the parameter $g_0$ in the QS model in Eq.~(\ref{eq:hamQSM}). The UM model exhibits a sharp transition at $\alpha_c=1/\sqrt{2}$ and the properties of the transition point can be tuned by $J_c$. However, here, we fix $N=1$, $J_c=1$ and vary $\alpha$ only.

\subsection{Rosenzweig-Porter model} \label{sec:def_RPmodel}

Finally we consider the Rosenzweig-Porter (RP) random matrix ensemble~\cite{Rosenzweig60} and its generalized form~\cite{Kravtsov15}.
We introduce the corresponding RP model in a form of a quadratic model given by Eq.~\eqref{eq:ham_ensemble}, for which the Hamiltonian matrix elements are given by
\begin{equation} \label{def_hij_RP}
h_{i,j}=h_{j,i}=\frac{\lambda\;\mu_{i,j}}{D^{c/2}}\;,
\end{equation}
{even though later we consider the RP model independently from other quadratic and interacting models.}
The parameters $\lambda$ and $c$ in Eq.~(\ref{def_hij_RP}) are real, and $\mu_{i,j}$ are independent and identically distributed random variables drawn from a box distribution, $\mu_{i,j} \in [-1,1]$$~$\cite{Bera18}.

Previous studies of the RP model reported emergence in three different regimes~\cite{Kravtsov15, Facoetti16, Truong16, Monthus17, vonSoosten19}: at $c<1$ the system is fully ergodic, at $1<c<2$ the eigenstates are fractal and the system is in a nonergodic extended regime, and at $c>2$ the eigenstates are localized in the computational basis. 
Numerical investigation of the short-range spectral statistics$~$\cite{Pino19} indicates that there is a change of the nearest-neighbour level statistics at $c=2$ from the GOE to the Poisson statistics.
We refer to this transition as the eigenstate transition in the RP model.
The statistics of the nearest-neighbour gaps at the transition can be tuned by the parameter $\lambda$.
We note that the system at the transition, $c=2$, is not multifractal but fully localized~\cite{Kravtsov15, Pino19}.
This is an important difference when compared to the PLRB model, which, as we argue in this paper, is also manifested in the dynamical properties of the RP model~\cite{Bera18, DeTomasi19}.

\section{Measures of quantum dynamics} \label{sec:methods}

\subsection{Spectral form factor (SFF)}

We study quantum dynamics of initial states evolved under the Hamiltonian $\hat H$. The Hamiltonian eigenspectrum is characterized by the SFF
\begin{equation}
\label{eq:sff_raw_one_real}
K^{H}(t)=\frac{1}{D^2}\left| \sum_{\nu=1}^D  e^{-iE_\nu t}  \right|^2 \;,
\end{equation}
where $E_\nu$ are eigenenergies of $\hat H$. The SFF is further averaged over different Hamiltonian realizations, denoted by $\langle ... \rangle_H$, which gives rise to
\begin{equation}
\label{eq:sff_raw}
K^{}(t)=\bigl< K^{H}(t) \bigl>_H\;.
\end{equation}
At long times, $K^{}(t)$ saturates to the corresponding long-time value
\begin{equation} \label{def_Kbar}
\overline{K} = \lim_{t\to\infty} K^{}(t) = \frac{1}{D}\;\;,
\end{equation}
which equals the time-averaged value over a sufficiently large time interval.
The SFF is a Fourier transform of energy level distances and it gives access to both short-range and long-range spectral correlations.
Here we argue that it may also represent a useful tool for understanding the behavior of survival probabilities in the quench dynamics studied below. 

\subsection{Survival probability}

We are interested in quantum quenches from the eigenstates $\{|m\rangle\}$ of the initial Hamiltonian $\hat H_0$ to the final Hamiltonian $\hat H$ with eigenstates $\{|\nu\rangle\}$. The survival probability of an eigenstate $|m\rangle$ is, for a given realization of the Hamiltonian $\hat H$, defined as 
\begin{equation}
\label{eq:sur_prob}
P_{m}^{H}(t)=|\langle m | e^{-i\hat Ht} |m \rangle  |^2 =
\bigg| \sum_{\nu=1}^D  |c_{\nu m}|^2  e^{-iE_\nu t}  \bigg|^2,
\end{equation}
where we set $\hbar \equiv 1$, $D$ is the Hilbert-space dimension, $c_{\nu m}=\langle \nu | m \rangle$ is the overlap of $|m\rangle$ with $| \nu \rangle$, and $E_\nu$ is an eigenenergy of $\hat H$.
The averaged survival probability is defined as
\begin{equation} \label{eq:sur_averaged}
P(t)=\langle\langle P_m^{H}(t)\rangle_m\rangle_{H},
\end{equation}
where $\langle ... \rangle_m$ denotes the average over {\it all} eigenstates $|m\rangle$ of the initial Hamiltonian $\hat H_0$, and $\langle ... \rangle_H$ denotes the average over different realizations of the final Hamiltonian $\hat H$.

At long times, $P(t)$ approaches $\overline{P}$, which is equal to the average inverse participation ratio of eigenstates of $\hat H$ in the eigenbasis of $\hat H_0$, $\overline{P_{}^{}}=\langle\langle \sum_\nu  |c_{\nu m}|^4 \rangle_m\rangle_{H}$.
We express $\overline{P}$ as
\begin{equation} \label{def_Pbar}
\overline{P} = \left\langle\left\langle \sum_{\nu=1}^D  |c_{\nu m}|^4 \right\rangle_m\right\rangle_{H} =  P_\infty + c_P D^{-\gamma} \;,
\end{equation}
i.e., as a sum of the nonzero asymptotic value
\begin{equation}
P_\infty=\lim_{D \rightarrow \infty} \overline{P_{}^{}} \;,
\end{equation}
and a part that vanishes in the thermodynamic limit $D \to \infty$ as $\propto D^{-\gamma}$, where $\gamma>0$ is the fractal dimension.
We distinguish between two scenarios that give rise to nonzero $P_\infty>0$: the case of a mobility edge, when a fraction of eigenstates $|m\rangle$ are localized in the eigenbasis spanned by $|\nu\rangle$ and the remaining fraction is not localized, and the case of complete localization when all eigenstates $|m\rangle$ are localized.
When using Eq.$~$\eqref{def_Pbar}, the definition of $\gamma$ as the fractal dimension is only meaningful in the former case.

\subsection{Relationship between the SFF and survival probability} \label{sec:relationship}

A common interpretation of the SFF~(\ref{eq:sff_raw_one_real}) is that it is only a measure of the Hamiltonian spectrum $\{E_\nu\}$.
An alternative perspective that we pursue here is that the SFF can be viewed as the survival probability~(\ref{eq:sur_prob}) of a special initial state $ |T\rangle$, namely the infinite-temperature pure state 
\begin{equation} \label{eq:def_T}
|T \rangle = \frac{1}{\sqrt{D}}\sum_{\nu=1}^D  |\nu \rangle \;,
\end{equation}
such that $K^H(t) \to P_T^H(t)$, and the averaging is then only performed over different Hamiltonian realizations.
This relationship is valid for any Hamiltonian $\hat H$ that dictates the dynamics.

The main goal of this paper is to study survival probabilities from different types of initial states, including the state~(\ref{eq:def_T}) that provides the link to the SFF, with the focus on the dynamics at criticality.
Specifically, the initial states are of three types.
(a) The one from Eq.~(\ref{eq:def_T}), for which the survival probability corresponds to the SFF.
(b) Product states $|m\rangle = |I\rangle$ in site occupation (computational) basis for quadratic models, and in Fock space for interacting models.
We refer to this type of states as site-localized (Fock-space-localized) states.
(c) Product states in quasimomentum occupation basis, $|m\rangle = |k\rangle = \frac{1}{\sqrt{D}}\sum_k e^{ikI}|I\rangle$, which correspond to conventional plane waves for quadratic models, and to plane waves in the Fock space for interacting models.
For brevity we refer to them as plane waves for all models under consideration.

We note that in the special case of the GOE matrices, there is an exact asymptotic relationship between the SFF and survival probabilities from other types of initial states such as site-localized (Fock-space-localized) states and plane waves.
We further elaborate on this relationship in Sec.~\ref{sec:GOE} in the context of Eq.~(\ref{eq:K_p_relation}).

The choice of the initial states imply the value of the time averaged $\overline{P}$ from Eq.~(\ref{def_Pbar}).
In the case (a), it follows from Eq.~(\ref{def_Kbar}) that $P_\infty=0$ and $\gamma=1$.
In the case (b) one gets $P_\infty = 0$ in the fully delocalized regime of $\hat H$, while $P_\infty > 0$ in the localized regime or in the regime with a mobility edge.
If the initial wave function at the transition exhibits (multi)fractal properties in the eigenbasis of $\hat H$, one expects $\gamma < 1$. 
In the case (c) one again gets $P_\infty = 0$ and $\gamma \approx 1$ for all cases under consideration here. We note, however, that the latter is not the case in the Aubry-André model~\cite{hopjan2023}.

We study survival probabilities in the units of scaled time $\tau$,
\begin{equation} \label{def_tau}
\tau = t/t_{H}^{\rm typ} \;,\;\;\;  t_{H}^{\rm typ}=2 \pi/\delta E^{\rm typ} \;,  
\end{equation}
where $t_{H}^{\rm typ}$ is the typical Heisenberg time, $\delta E^{\rm typ}$ is the typical level spacing, $\delta E^{\rm typ}= \exp[\langle \langle \ln( E_{\nu+1} -E_\nu) \rangle_{\nu} \rangle_{H}]$, and $\langle ... \rangle_\nu$ denotes the average over all pairs of nearest levels.
We note that in Hamiltonian systems, the SFF is commonly studied after spectral unfolding that removes the impact of the density of states.
However, if one seeks to establish the connection of the SFF with the survival probabilities, which are measured either as a function of raw time $t$ or the scaled time $\tau$, no unfolding is carried out.
In Appendix~\ref{app:1} we compare the SFF with and without unfolding, and we show that the main features of our results are independent of this choice.

\section{scale-invariant dynamics} \label{sec:scaleinvariant}

As discussed in introduction, a well-established universality of the SFF occurs in the quantum chaotic regime, when the SFF follows the so-called ramp, predicted by the RMT~\cite{mehta_91,Santos2017, torresherrera_garciagarcia_18, Schiulaz_19}.
Specifically, the quantity that actually exhibits the scale-invariant (i.e., $L$-independent) behavior in the quantum chaotic regime is the normalized SFF $k(\tau)$,
\begin{equation} \label{def_ktau}
    k(\tau) = \frac{K(\tau)}{\overline{K}} \;,
\end{equation}
such that at long times one gets $k(\tau\gg 1) = 1$.
The normalized $k(\tau)$ from Eq.~(\ref{def_ktau}) has been applied in recent studies of quantum chaos in interacting systems, see, e.g.,~Refs.~\cite{suntajs_bonca_20a, Sierant2020, suntajs_prosen_21, Prakash21, suntajs_vidmar_22, hopjan2023, Fritzsch23a}. For brevity we refer to the normalized SFF as the SFF.

In what follows, our main goal is to explore whether a similar form of scale-invariant behavior can also be observed at eigenstate transitions from quantum chaos to localization.

\subsection{From SFF to survival probability}

Inspired by the scale-invariant behavior of the SFF in the quantum chaotic regime~(\ref{def_ktau}), and the relationship between the SFF and survival probabilities from Sec.~\ref{sec:relationship}, we now explore features of scale invariance in survival probabilities.
In Ref.~\cite{hopjan2023} we introduced the scaled survival probability $p(\tau)$, henceforth survival probability,
\begin{equation}
\label{eq:sur_prob_norm}
p(\tau)= \frac{P(\tau)-P_\infty}{\overline{P_{}^{}}-P_\infty} \;,
\end{equation}
which should be considered as an analog to the SFF defined in Eq.~(\ref{def_ktau}).
In particular, the survival probability saturates at long times to $p(\tau\gg 1) = 1$, and the subtraction with $P_\infty$ is used to make sure that in the case of mobility edges in the spectrum, $P(\tau)$ decays to zero in the thermodynamic limit $D \to \infty$.
As discussed above, for certain initial states such as those that make the connection to the SFF, one gets $P_\infty = 0$ and hence no subtraction is needed in Eq.~(\ref{def_ktau}).
Also, we will show in Sec.~\ref{sec:RPmodel} that at the eigenstate transition in the RP model, at which the eigenstates are fully localized and hence $\overline{P}=P_\infty$ for site-localized initial states in finite systems, again no subtraction is needed in Eq.~(\ref{eq:sur_prob_norm}).

Below we illustrate scale invariance of survival probabilities for the studied initial states in the case of GOE Hamiltonians, and in the remainder of the paper, see Secs.~\ref{sec:Anderson_PLRB}-\ref{sec:RPmodel}, we numerically test scale-invariant properties at criticality for various other models.

\subsection{Scale invariance of GOE Hamiltonians} \label{sec:GOE}

The case in which the Hamiltonian is represented by a GOE matrix serves as a benchmark for quantum chaotic dynamics as it allows for establishing analytical predictions~\cite{mehta_91}.
Here we provide exact expressions for survival probabilities from all the initial states studied in this paper.
More generally, these results will also serve as a motivation in the subsequent sections in the search for scale-invariant dynamics at criticality.

\begin{figure}[!t]
\centering
\includegraphics[width=0.98\columnwidth]{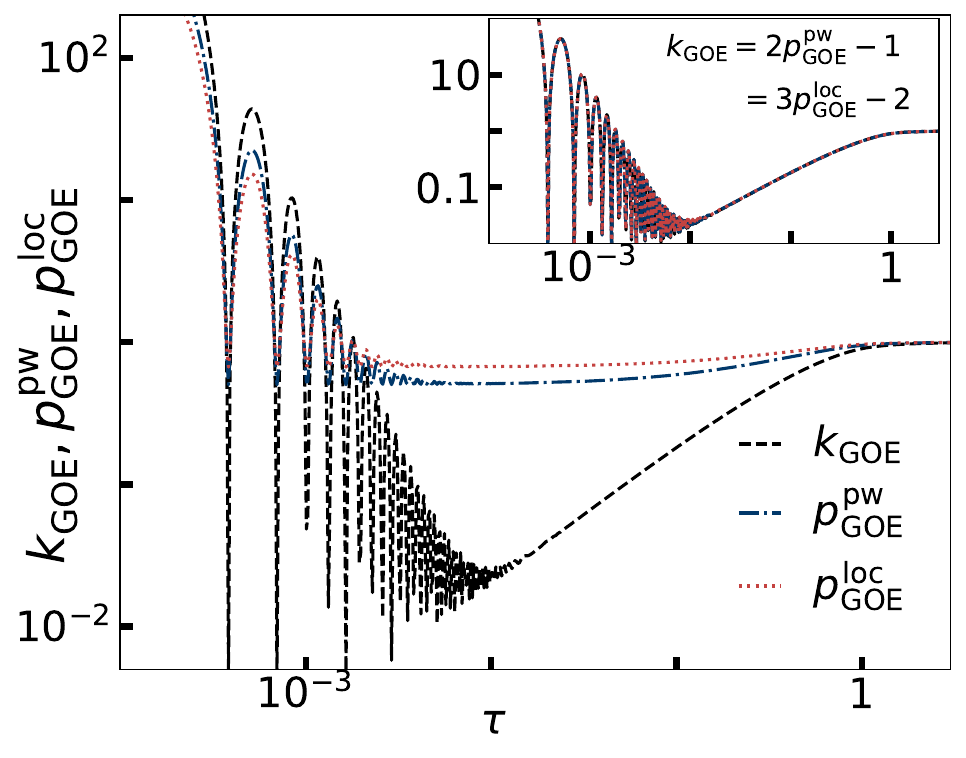}
\caption{
The GOE results for $p_{\rm GOE}^{}(\tau;c_p)$ from Eq.~\eqref{eq:p_goe} at $D=2500$.
Main panel: $k_{\rm GOE}^{}(\tau)$, $p_{\rm GOE}^{\rm loc}(\tau)$ and $p_{\rm GOE}^{\rm pw}(\tau)$, see text for definitions.
(Inset) $k_{\rm GOE}^{}(\tau)$, $3p_{\rm GOE}^{\rm loc}(\tau)-2$ and $2p_{\rm GOE}^{\rm pw}(\tau)-1$.
The curves exhibit a perfect collapse, as expected from Eq.~(\ref{eq:K_p_relation}).
}
\label{fig_2}
\end{figure}

For concreteness, we define the GOE Hamiltonian in analogy to Eq.~(\ref{eq:ham_ensemble}), where the matrix elements $h_{i,j}$ are i.i.d.~random numbers drawn from a Gaussian distribution with zero mean and variance 2 (1) for the diagonal (off-diagonal) matrix elements.
Analytical predictions for the survival probability from initially site-localized states, evolving under the GOE Hamiltonian, were derived in~\cite{Santos2017, torresherrera_garciagarcia_18, Schiulaz_19}.
While Refs.$~$\cite{Santos2017,torresherrera_garciagarcia_18,Schiulaz_19} focused on the particular initial state, we here adapt the formula therein to describe the survival probabilities of all initial states under consideration.
We first express time in units of the corresponding Heisenberg time $\tau=t/t_H^{\rm typ}$ from Eq.~(\ref{def_tau}), which is $t_{H}^{\rm typ}=2\sqrt{D}$ for the GOE Hamiltonian.
In this units, Eq.~(8) from Ref.~\cite{Schiulaz_19} can be written as
\begin{equation}
\label{eq:P_GOE}
P_{\rm GOE}^{}(\tau;\overline{P})= \frac{1-\overline{P}}{D-1}\Biggl[D \frac{J_{1}^{2}(4D\tau)}{(2D\tau)^{2}}-b_2(\tau)\Biggr]+\overline{P}\;,
\end{equation}
where  $J_{1}^{}$ is the Bessel function of the first kind, and the two-point function $b_2(\tau)$~\cite{mehta_91} is given by
\begin{align}
b_2(\tau)&=[1-2\tau+\tau\ln(1+2\tau)]\Theta(1-\tau) \nonumber \\ 
&+\{\tau\ln[(2\tau+1)/(2\tau-1)]-1\}\Theta(\tau-1) \;.
\end{align}
The first term on the right-hand side~of Eq.~(\ref{eq:P_GOE}) quickly vanishes at $\tau>1$ and hence the limit $P_{\rm GOE}\to\overline{P}$ is reached at long times $\tau>1$. 

\begin{figure}[!t]
\centering
\includegraphics[width=0.98\columnwidth]{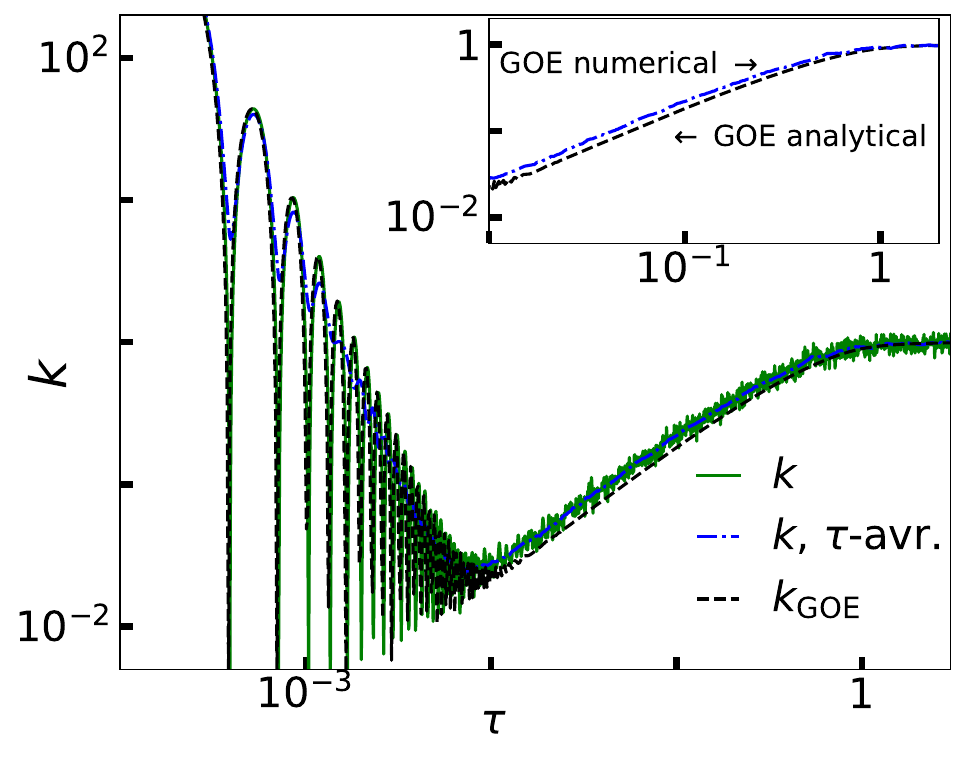}
\caption{Numerical results of the SFF $k^{}(\tau)$ from Eq.$~$\eqref{def_ktau} (green) compared to the analytical prediction for the SFF $k_{\rm GOE}^{}(\tau)$ from Eq.$~$\eqref{eq:kgoe_pgoe} (dashed black) at $D=2500$.
We additionally perform the running average to obtain $\tau$-averaged $k^{}(\tau)$ (dashed-dotted blue).
(Inset) Details in the comparison of the GOE ramp from Eq.$~$\eqref{eq:K_goe_ramp} to the numerical results that show slight differences in the slope of the ramp.}
\label{fig_3}
\end{figure}

For the GOE Hamiltonian, the time-averaged survival probabilities $\overline{P}$~(\ref{def_Pbar}) have $P_\infty = 0$ and $\gamma=1$ for all initial states of interest.  Hence, the dependence of $P$ on $\overline{P}$ can be expressed as the dependence on the coefficient $c_P$. 
{For the site-localized states one expects $\overline{P}=3/D$ \cite{Santos2017, torresherrera_garciagarcia_18, Schiulaz_19}, while for the infinite-temperature state in Eq.~(\ref{eq:def_T}) [i.e., the SFF] one expects $\overline{P}=1/D$ and for the plane waves one expects $\overline{P}=2/D$. 
According to Eq.~(\ref{def_Pbar}), this gives: $c_P=3$ for the site-localized states, $c_P=1$ for the initial state in Eq.~(\ref{eq:def_T}),}  and $c_P=2$ for the plane waves.
Then, following Eq.$~$\eqref{eq:sur_prob_norm} we rescale the survival probability from Eq.~(\ref{eq:P_GOE}) to obtain
\begin{equation}
\label{eq:p_goe}
p_{\rm GOE}^{}(\tau;c_P)=\frac{1}{c_P} \Biggl[D \frac{J_{1}^{2}(4D\tau)}{(2D\tau)^{2}}-b_2(\tau)\Biggr]+1 \;,
\end{equation}
where we assumed $(1-c_P/D)/(1-1/D) \to 1$.

The term in Eq.~(\ref{eq:p_goe}) that contains the Bessel function is the only term that contains the information about the Hilbert space $D$, however, it quickly decays to zero since its envelope function is given by $1/(c_p 8\pi D^2 \tau^3)$.
It is then clear from Eq.~(\ref{eq:p_goe}) that at sufficiently long times $p_{\rm GOE}$ becomes scale-invariant, i.e., it only depends on $\tau$ for all system sizes.
The crossover time to the scale-invariant dynamics is nowadays known as the Thouless time, which in the case of GOE Hamiltonians scales as $\tau^{\rm Th}_{\rm GOE} = (3/\pi)^{1/4}/(2 \sqrt{D})$.
Scale invariance of $p_{\rm GOE}(\tau)$ in the GOE case can be understood as the extension of the scale invariance of the SFF $k_{\rm GOE}(\tau)$ from Eq.~(\ref{def_ktau}).
Specifically, Eq.~(\ref{eq:p_goe}) suggest that 
\begin{equation} \label{eq:kgoe_pgoe}
    k_{\rm GOE}(\tau) = c_p \, p_{\rm GOE}(\tau; c_p) + (1-c_p)\;,
\end{equation}
and hence $k_{\rm GOE}^{}(\tau)=p_{\rm GOE}^{}(\tau; c_P=1)$, giving rise to the well-known expression for the GOE ramp in the SFF,
\begin{equation}
\label{eq:K_goe_ramp}
k_{\rm GOE}^{}(\tau_{\rm GOE}^{Th}<\tau\lesssim 1)\approx1-b_2(\tau) = 2\tau - \tau \ln(1+2\tau) \;.
\end{equation}
Beyond the emergence of scale invariance at $\tau > \tau_{GOE}^{Th}$ it is also interesting to consider the relationships between different survival probabilities, which are valid for arbitrary time $\tau$.
Let us define $p_{\rm GOE}^{\rm loc}(\tau) = p_{\rm GOE}(\tau;c_p=3)$ for the initial site-localized states and $p_{\rm GOE}^{\rm pw}(\tau) = p_{\rm GOE}(\tau;c_p=2)$ for the initial plane waves.
Equation~(\ref{eq:kgoe_pgoe}) then implies
\begin{equation}
\label{eq:K_p_relation}
k_{\rm GOE}^{}(\tau)=2p_{\rm GOE}^{\rm pw}(\tau)-1=3p_{\rm GOE}^{\rm loc}(\tau)-2\;.
\end{equation}
This relation will also be tested in the dynamics at criticality in the next sections. 

The results of the above discussion are illustrated in Fig.~\ref{fig_2} at $D=2500$.
In the main panel of Fig.~\ref{fig_2} we show $k_{\rm GOE}^{}(\tau)$, $p_{\rm GOE}^{\rm loc}(\tau)$, and $p_{\rm GOE}^{\rm pw}(\tau)$.
The SFF $k_{\rm GOE}^{}(\tau)$ exhibits a sharp decay until it hits the GOE ramp at the Thouless time $\tau_{\rm Th}^{\rm GOE}$.
In contrast, the other survival probabilities exhibit a broad minimum, which is also referred to as the correlation hole~\cite{Schiulaz_19}.
From these results only, one may not be able to exclude the possibility that the Thouless time $\tau_{\rm Th}^{\rm GOE}$ depends on the specific type of initial states.
However, when one plots the rescaled survival probabilities $3p_{\rm GOE}^{\rm loc}(\tau)-2$ and $2p_{\rm GOE}^{\rm pw}(\tau)-1$ from Eq.~(\ref{eq:K_p_relation}), see the inset of Fig.~\ref{fig_2}, one observes a perfect collapse of all survival probabilities suggesting that, as expected, $\tau_{\rm Th}^{\rm GOE}$ is independent of the choice of the initial state.

While the results in Fig.~\ref{fig_2} are obtained from the analytical expressions, we also test them numerically and compute $k^{}(\tau)$, $p_{}^{\rm pw}(\tau)$ and $p_{}^{\rm loc}(\tau)$ of the GOE Hamiltonians.
We first verified (not shown) that the relationship between these quantities, given by Eq.~\eqref{eq:K_p_relation}, also holds for the numerical results.
Thus it is sufficient to only focus on the SFF $k^{}(\tau)$.
Figure~\ref{fig_3} compares the numerical and analytical results $k(\tau)$ and $k_{\rm GOE}(\tau)$, respectively.
The numerical results follow the behavior of the analytical result, and we smoothen the noise by carrying out the running averages in time.
The inset of Fig.~\ref{fig_3}, however, highlights that the numerical result differs from the analytical result, since the numerical result lies higher than the analytical one and it follows a slightly different slope of the ramp.
We attribute this deviation to the lack of spectral unfolding and filtering in the numerical results for the SFF. 
In Appendix~\ref{app:1}, see Fig.~\ref{fig_20}, we compare the SFFs with and without unfolding and filtering, and we show that in the former case the numerical results fit the analytical predictions. 
In the reminder of the paper, when the results are compared to the GOE predictions, they are always compared to the numerical results for the GOE Hamiltonians.
In Appendix~\ref{app:2} we also provide further details about the numerical averaging over Hamiltonian realizations and the extraction of the Thouless time $\tau_{\rm Th}$.

\subsection{Arguments for scale invariance at criticality} \label{sec:arguments}

We now turn our attention to criticality and give arguments for the emergence of scale invariance in quadratic models.
Perhaps one of the earliest papers addressing scaling properties at criticality is Ref.~\cite{Chalker88}, where the two-particle spectral function was investigated.
The study revealed a novel scaling form of the two-particle spectral function emerging at the critical point (and coexisting with diffusive behavior), which was conjectured to be a consequence of the fractal structure of eigenstates at criticality.
The ansatz of Ref.~\cite{Chalker88} was then argued to be consistent with the emergence of power-law behavior seen in the survival probabilities from initially localized states, i.e., $P_{}^{\rm loc}(t)\propto t^{-\beta}$~\cite{Huckestein94,Brandes96,Ketzmerick_97,Kravtsov10,Kravtsov11}.
The studies confirmed that the power-law exponent $\beta$ is related to the multifractal dimension of the eigenstates at criticality~\cite{Huckestein94, Brandes96, Ketzmerick_97, Kravtsov10, Kravtsov11} even though they did not investigate systematically the system size dependence of $P_{}^{\rm loc}(t)$.
While the ansatz of Ref.~\cite{Chalker88} predicts scale-invariant properties, we note that it is based on properties of eigenstates a within microcanonical window around critical states.
General initial states may ultimately be projected to all eigenstates and the scale invariance might be hindered in $P_{}^{\rm loc}(t)$ by the effect of mobility edges.

Our ansatz for the scaled survival probability, introduced in Ref.~\cite{hopjan2023} and in Eq.~(\ref{eq:sur_prob_norm}) of this paper, generalizes the above considerations.
For the initial site-localized states it predicts the scale-invariant power-law decay at criticality,
\begin{equation} \label{def_taubeta}
    p^{\rm loc}(\tau) = a_0\, \tau^{-\beta} \;,
\end{equation}
where $a_0$ and $\beta$ are fitting parameters. 
In particular, the ansatz from Eq.~(\ref{eq:sur_prob_norm}) includes two refinements:
First, the survival probability is subtracted by $P_\infty$, which is necessary in order to observe scale invariance in models with mobility edges, such as the Anderson models, and second, time is measured in units of the typical Heisenberg time $t_H^{\rm typ}$.
The later is crucial in models such as the Aubry-André model~\cite{hopjan2023}, in which the typical and the average Heisenberg time scale differently.
These refinements give rise to two important advantages of the ansatz:
first, the scale invariance is not only present in the power-law regime in the mid-time dynamics, but also in the late-time dynamics (as in the SFF), and second, it allows for a generalized relationship between the fractal dimension $\gamma$ and the power-law exponent $\beta$,
\begin{equation} \label{eq:def_gamma_beta}
\gamma = n \beta\;,\;\;\; {\rm with}\;\;\; t_H^{\rm typ} = D^n\;,
\end{equation}
where $n$ determines the scaling of the typical Heisenberg time with the Hilbert-space dimension $D$.  We note that the ansatz of scale-invariant power-law decay as in Eq.$~$\eqref{def_taubeta} seems to work also for driven critical systems$~$\cite{Chen23}. 

We here add two remarks.
The first is that that the scale-invariant power-law decay of $p^{\rm loc}(\tau)$ in Eq.$~$\eqref{def_taubeta} implies scale invariance of the power-law decay of the subtracted, but unscaled survival probability $P_{}^{\rm loc}(t)-P_{\infty}$ in the mid-time dynamics.
This statement is illustrated by rewriting Eqs.~(\ref{eq:sur_prob_norm}) and~(\ref{def_Pbar}) as
\begin{equation}
    P^{\rm loc}(t) - P_\infty = c_P D^{-\gamma} a_0 t^{-\beta} (t_H^{\rm typ})^\beta \;,
\end{equation}
which, using the relationship from Eq.~(\ref{eq:def_gamma_beta}), yields
\begin{equation}
    P_{}^{\rm loc}(t)-P_{\infty} = c_P a_0 t^{-\beta}\;.
\end{equation}
The second remark is that the reverse of the first remark is not true.
This can be illustrated by the power-law decay $P_{}^{\rm loc}(t)-P_\infty\propto t^{-\beta}$, which may emerge in the localized regime in finite systems~\cite{Brandes96, hopjan2023}.
However, the latter do not imply scale invariance of $p^{\rm loc}(\tau)$ in the mid-time dynamics~\cite{hopjan2023}.
Here, the dynamics is dominated by the projection to localized eigenstates, for which the ansatz from Ref.~\cite{Chalker88} is lost. 

A seemingly different, but not unrelated aspect is the scale invariance of the SFF at criticality.
Early study addressed the level statistics at criticality focusing on short-range statistics~\cite{Shklovskii93}, which was shown to be neither of Wigner surmise nor of Poisson statistics.
The long-range spectral correlation~\cite{Kravtsov94, Aronov95}, measured by spectral rigidity, were further argued to be related to the eigenfunction correlations at criticality~\cite{Chalker96a,Chalker96b,Mirlin00}.
Remarkably, in Ref.$~$\cite{Chalker96b} the connection between the power-law decay of $P_{}^{\rm loc}(t)$ and the plateau in the SFF $K(t)$ was conjectured.
Then, it may not be unexpected, at least for quadratic models, that the scale-invariant power-law decay of $p_{}^{\rm loc}(\tau)$ is accompanied by the scale-invariant plateau in the SFF $k(\tau)$.
Recent studies~\cite{suntajs_prosen_21, suntajs_vidmar_22} investigated system-size dependence of the plateau in the SFF $k(\tau)$ and indeed observed a scale-invariant behavior.


\section{Results for quadratic models} \label{sec:Anderson_PLRB}

\subsection{From single-particle quantum chaos to localization} \label{sec:Similarity}

We start our analysis of the dynamics with two quadratic models, the 3D Anderson model~(\ref{eq:ham3DA}) and the PLRB model~(\ref{eq:ham_ensemble}).
We first study the dynamics across the critical point at a fixed system size $L$,
and we focus on the unscaled SFF $K(\tau)$ from Eq.~(\ref{eq:sff_raw}) and the survival probability $P(\tau)$ from Eq.~(\ref{eq:sur_averaged}).

\begin{figure}[t]
\centering
\includegraphics[width=0.91\columnwidth]{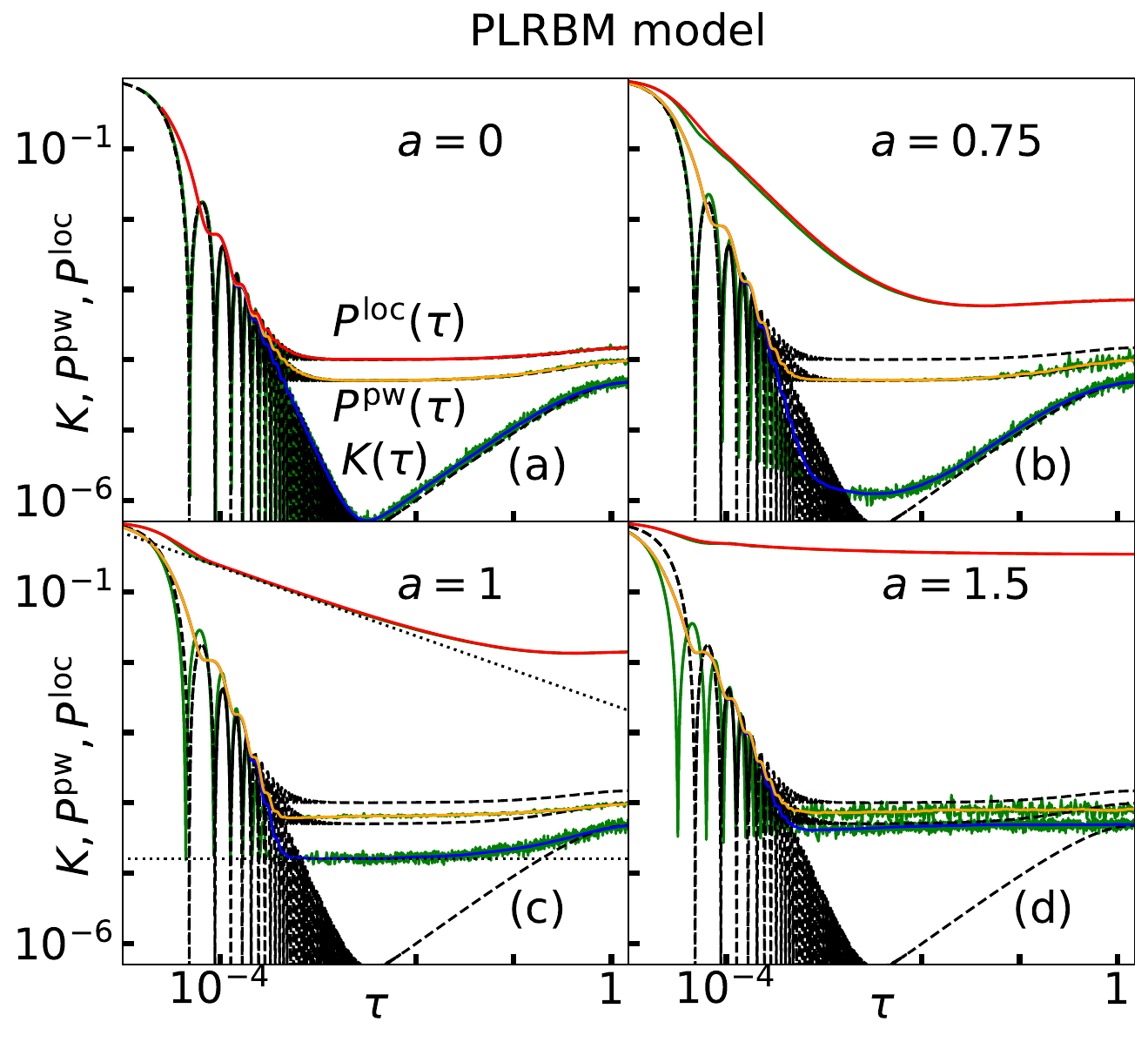}
\caption{
$K(\tau)$, $P^{\rm loc}(\tau)$ and $P^{\rm pw}(\tau)$ in the PLRB model at $b=0.5$ and $D=L=$~20 000.
(a) $a=0$, (b) $a=0.75$, (c) the critical point $a=1$, and (d) $a=1.5$. Green lines are results before time averaging.
The corresponding results for GOE Hamiltonians from Sec.~\ref{sec:GOE} (see Fig.~\ref{fig_2}) are shown as dashed black lines. 
The dotted lines in (c) indicate the power-law decay for $P_{}^{\rm loc}(\tau)$ and the plateau in $K(\tau)$. 
}\label{fig_4}
\end{figure}

Figure~\ref{fig_4} shows $K(\tau)$ and $P(\tau)$ in the PLRB model at different values of the parameter $a$ while keeping the parameter $b$ fixed.
The limit of $a=0$ is, up to normalization, the limit of the GOE Hamiltonians, and the numerical simulations fit the corresponding analytical predictions rather well, see Fig.$~$\ref{fig_4}(a).
Increasing the value of $a$ but keeping it below the transition, see Fig.$~$\ref{fig_4}(b), produces the most significant effect to $P^{\rm loc}(\tau)$ while $K(\tau)$ and $P^{\rm pw}(\tau)$ change only mildly.
At the transition point $a=1$, see Fig.$~$\ref{fig_4}(c), $P^{\rm loc}(\tau)$ develops a power-law decay in the mid-time dynamics and $K(\tau)$ develops a plateau in approximately the same time regime.
We also note a change in $P^{\rm pw}(\tau)$, which starts deviating from the analytical result at $a=1$.
In the localized regime, see Fig.$~$\ref{fig_4}(d), the deviations from the analytical results further increase.  

\begin{figure}[b!]
\centering
\includegraphics[width=0.91\columnwidth]{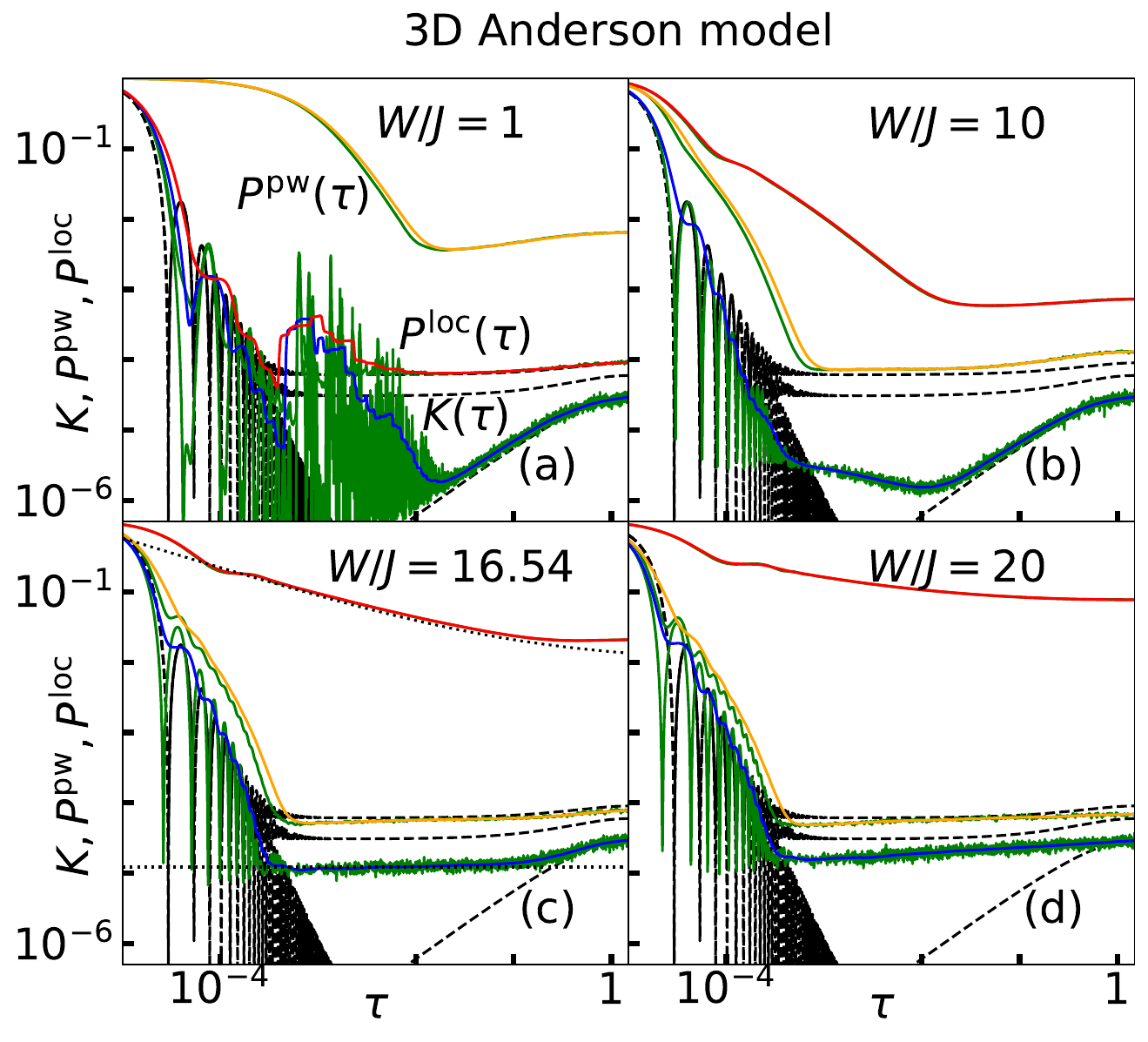}
\caption{
$K(\tau)$, $P^{\rm loc}(\tau)$ and $P^{\rm pw}(\tau)$ in the 3D Anderson model at $D=L^3=$~32 768.
(a) $W/J=1$, (b) $W/J=10$, (c) the critical point $W/J=16.54$, and (d) $W/J=20$. Green lines are results before time averaging.
The corresponding results for GOE Hamiltonians from Sec.~\ref{sec:GOE} (see Fig.~\ref{fig_2}) are shown as dashed black lines. 
The dotted lines in (c) indicate the power-law decay for $P_{}^{\rm loc}(\tau)$ to a nonzero constant and the plateau in $K(\tau)$.
}\label{fig_5}
\end{figure}

In Fig.~\ref{fig_5} we compare the dynamics in the PLRB model to those in the Anderson model, in which we vary the disorder $W$.
At small $W/J=1$ the long-time limit of $P^{\rm pw}(\tau)$ does not approach the limit of GOE but has a higher value, see Fig.$~$\ref{fig_5}(a).
This can be interpreted as a proximity to the translationally invariant limit where the single-particle eigenstates are plane waves. 
Increasing the value of $W$ to $W/J=10$ within the chaotic regime, see Fig.$~$\ref{fig_5}(b), the influence of the translationally invariant limit is weaker and the behavior is similar to the delocalized site of the PLRB model, compare Fig.$~$\ref{fig_4}(b) and Fig.$~$\ref{fig_5}(b).
At the transition point $W_c$, see Fig.$~$\ref{fig_5}(c), there is even higher similarity with the PLRB model, compare Fig.$~$\ref{fig_4}(c) and Fig.$~$\ref{fig_5}(c).
$P^{\rm loc}(\tau)$ develops a power-law decay in the mid-time dynamics and $K(\tau)$ develops a plateau in approximately the same time regime.
The situation is similar in the localized regime, see Fig.$~$\ref{fig_5}(d).
To summarize, while the small $a$ and $W$ limits of both models are distinct, their behavior close to criticality is almost identical.

\subsection{Nearest level spacing statistics} \label{sec:PLRB}

\begin{figure}[!b]
\centering
\includegraphics[width=\columnwidth]{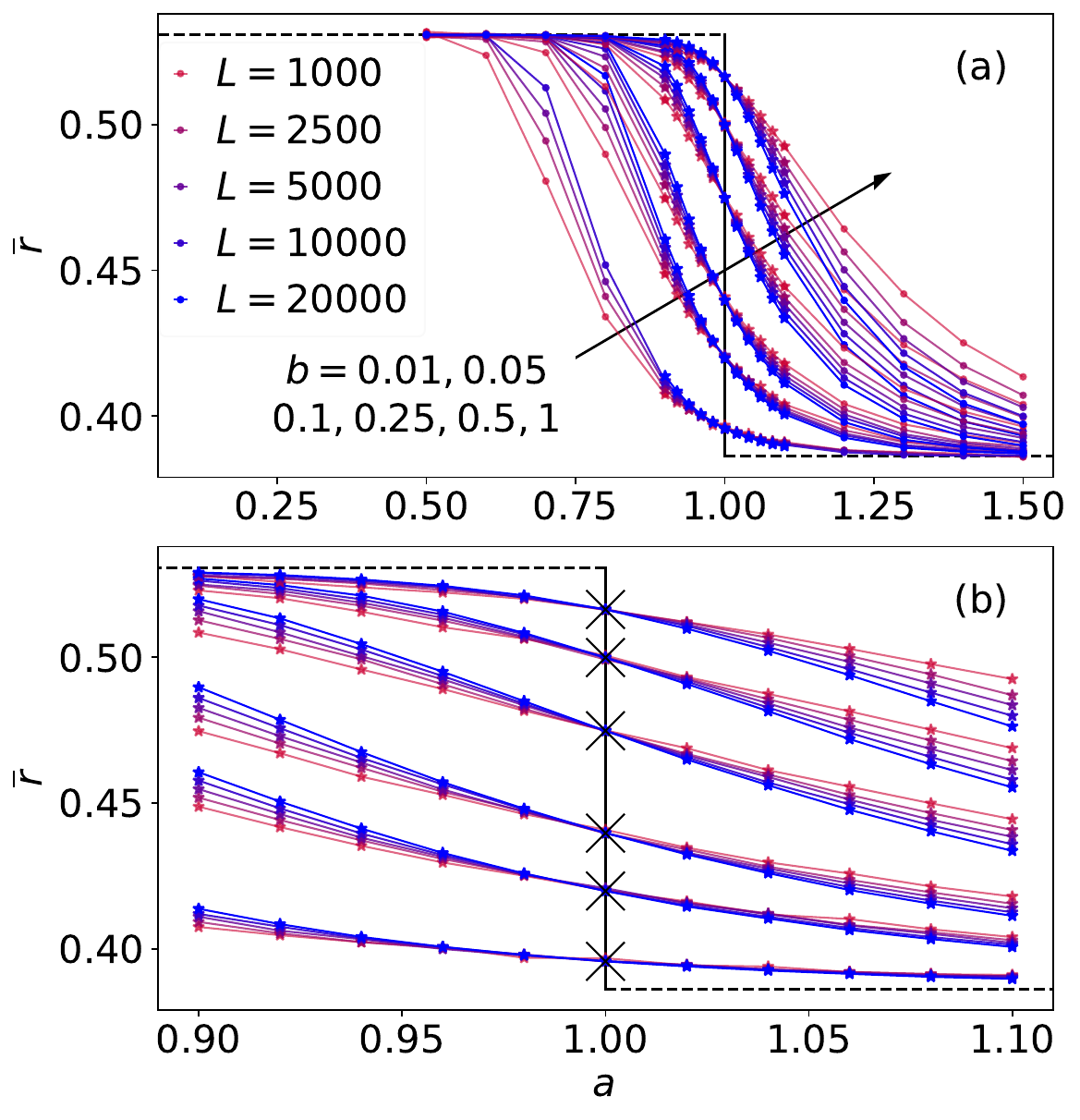}
\caption{
Average gap ratio $\bar r$ in the PLRB model as a function of $a$, for several values of $b$ and $L$.
The results in (b) are identical to those in (a), but zoomed into a narrow window around the critical point $a=1$.
The scale-invariant values of $\overline{r}$ at the critical point in (b) are denoted by the crosses (from bottom to top: $b= 0.01, 0.05, 0.1, 0.25, 0.5, 1$).
}\label{fig_6}
\end{figure}

\begin{figure}[!t]
\centering
\includegraphics[width=\columnwidth]{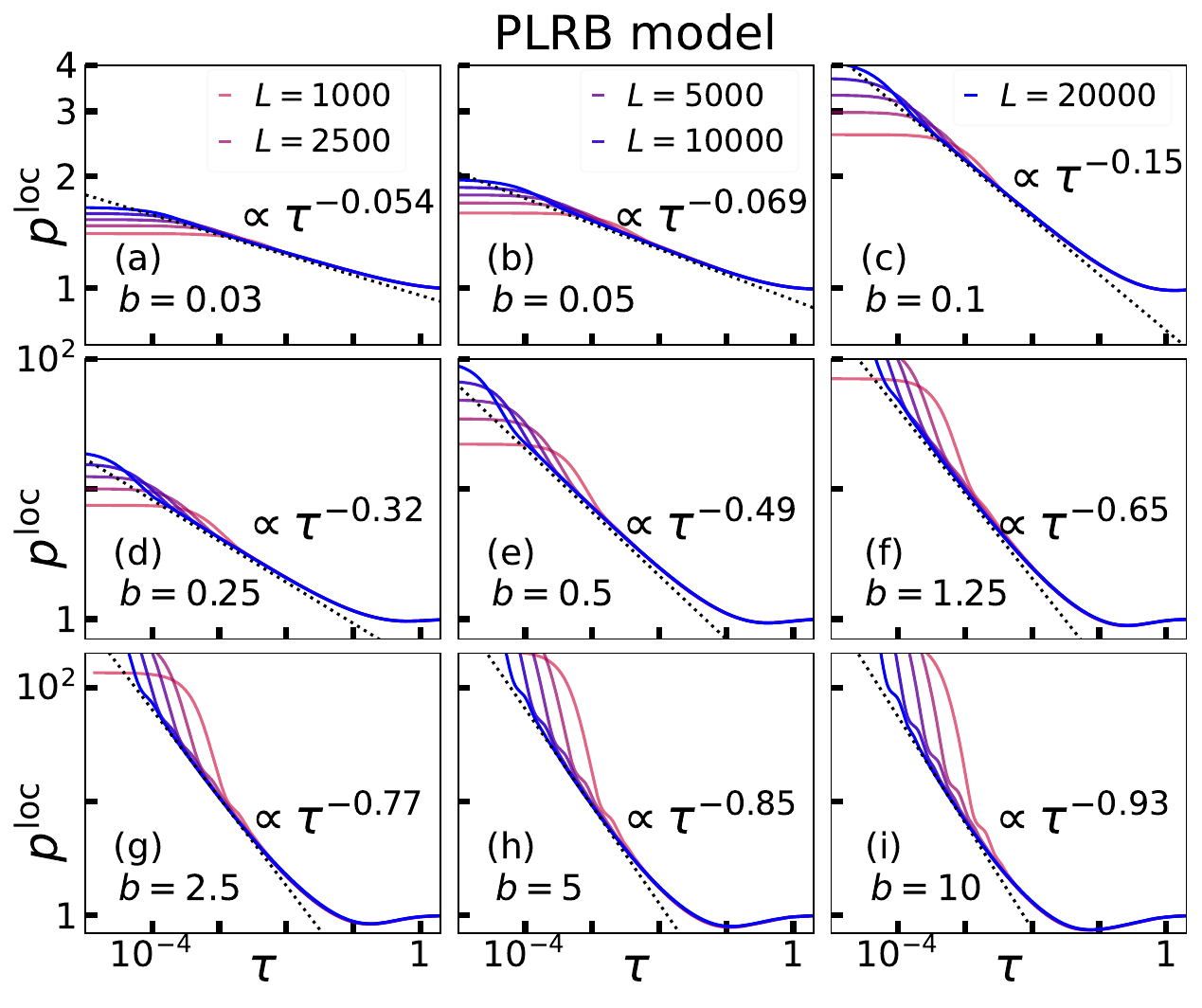}
\caption{
Survival probability $p_{}^{\rm loc}(\tau)$ from Eq.~\eqref{eq:sur_prob_norm} in the PLRB model at the critical point $a=1$, for different system sizes $L$.
Different panels  correspond to different values of the parameter $b$, ranging from small $b=0.03$ in (a) to large $b=10$ in (i).
The dotted lines are the fits $\propto \tau^{-\beta}$ from Eq.$~$(\ref{def_taubeta}), and in each panel we provide the value of $\beta$ obtained from the fit.
}\label{fig_7}
\end{figure}

\begin{figure*}[!b]
\centering
\includegraphics[width=2\columnwidth]{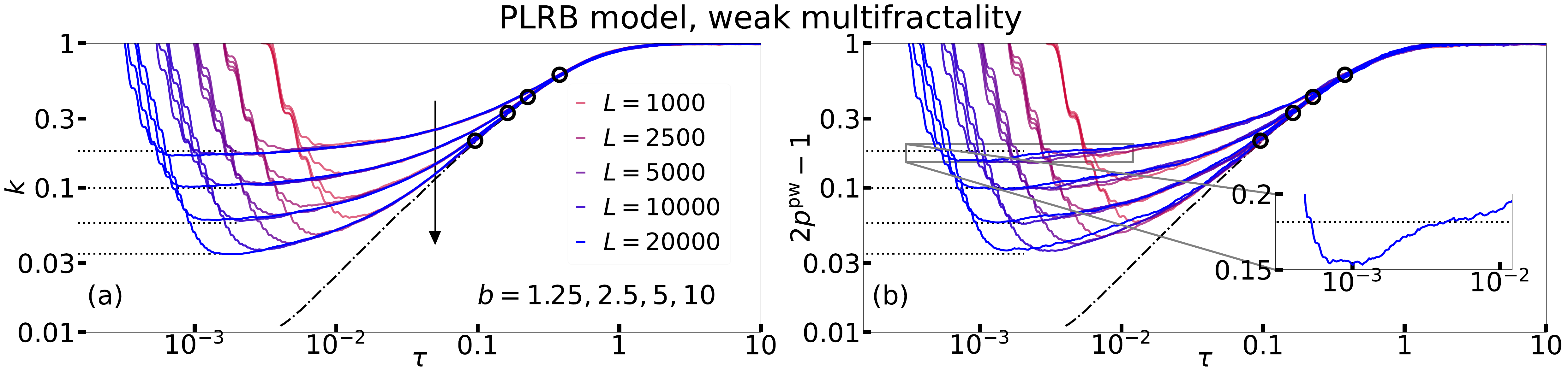}
\caption{
(a) The SFF $k(\tau)$, and (b) the survival probability from the initial plane waves $2p_{}^{\rm pw}(\tau)-1$, in the PLRB model at criticality, $a=1$.
Results are shown for several values of $b>1$ at weak multifractality, and different system sizes $L$, as indicated in the legend [the same parameters as in Figs.~\ref{fig_7}(f)--\ref{fig_7}(i)].
The horizontal dotted lines denote the plateau values $k_{\chi}^{}$, and the dashed-dotted lines denote the linear ramp of the numerically evaluated GOE Hamiltonians, cf.~Fig.$~$\ref{fig_3}.
The circles correspond to the extracted Thouless times $\tau_{\rm Th}$, at which $k^{}(\tau)$ approach the numerically evaluated GOE values. 
The inset in (b) shows a detail of the additional short-time feature of $2p_{}^{\rm pw}(\tau)-1$ compared to $k^{}(\tau)$.
}\label{fig_8}
\end{figure*}

We now focus on the scale-invariant properties at criticality.
We first demonstrate them for short-range spectral statistics, i.e., we calculate the ratio of consecutive level spacings of the single-particle spectrum,
\begin{equation}
r_\nu=\frac{\min\{\delta E_{\nu+1} ,\delta E_\nu\}}{\max\{\delta E_{\nu+1} ,\delta E_\nu\}}
\end{equation}
where $\delta E_{\nu} = E_{\nu+1} -E_\nu$ is the nearest level spacing. We define the average nearest level spacing ratio (shortly, the average gap ratio) as $\bar r = \langle\langle r_\nu \rangle_{\nu} \rangle_{H}$, with $\langle ... \rangle_\nu$ denoting the average over all pairs of nearest level spacings and $\langle ... \rangle_H$ denoting the average over 100 Hamiltonian realizations. 

While the change of the parameters $a$ and $W$ determines how close the systems are to the critical point, the parameters $b$ (in case of the PLRB model) and the dimensionality $d$ of the hypercubic lattice (in case of the Anderson models) are expected to change the spectral properties and the eigenstate properties at the critical point$~$\cite{Evers2000,Mirlin00}.
While results for the average gap ratios $\bar r$ of the Anderson models were reported elsewhere (see, e.g., Refs.~\cite{Tarquini17,suntajs_prosen_21}), we here focus on the PLRB models.

The results for $\bar r$ as a function of $a$ are shown in Fig.$~$\ref{fig_6} for several values of $b$ and $L$.
For each value of $b$ there exist a scale-invariant value $\bar r$ at the critical point $a=1$, see the crosses in Fig.$~$\ref{fig_6}(b). 
By increasing $b$ the scale-invariant value of $\bar r$ increases from the Poisson-like value $r_{\rm P} = 2\ln 2 - 1 \approx 0.3863$$~$\cite{oganesyan_huse_07} to the GOE-like value $r_{\rm GOE}\approx 0.5307$$~$\cite{atas_bogomolny_13}. 
The dependence of $\bar r$ on $b$ is summarized in Fig.$~$\ref{fig_12}(b) (see below).
The change of the spectral properties at criticality is expected to be accompanied with the change of the multifractality of states~\cite{Mirlin96}.
In particular, the GOE-like value of $\bar r$ is associated with weak multifractality where $\gamma \approx 1$, and the Poisson-like value of $\bar r$ is associated with strong multifractality where $\gamma \approx 0$.
We next study an alternative perspective on the degree of multifractality of the critical wavefunctions, obtained from the mid-time dynamics.

\subsection{Critical dynamics in the PLRB model}

\begin{figure*}[!t]
\centering
\includegraphics[width=2\columnwidth]{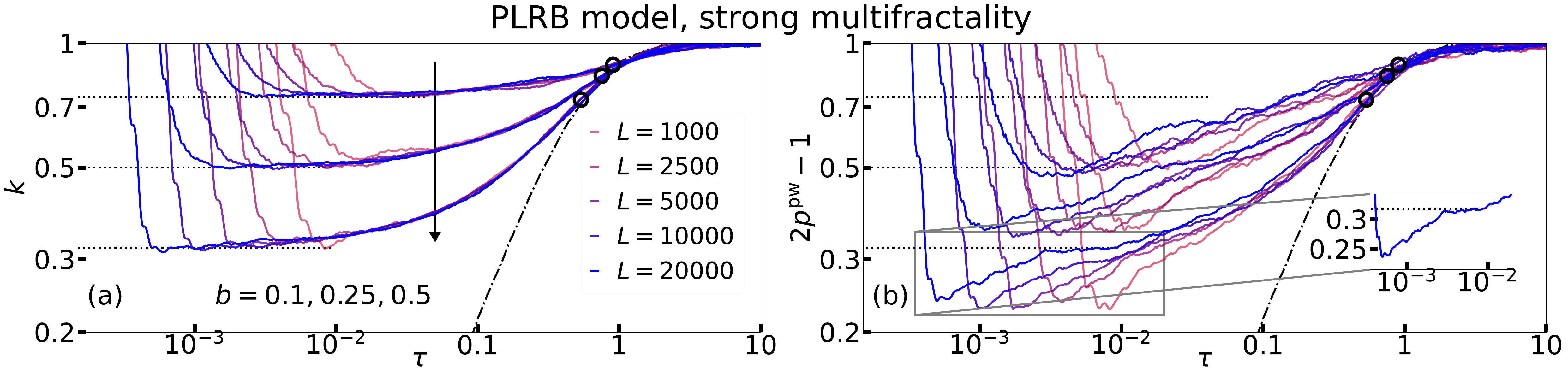}
\caption{
(a) The SFF $k(\tau)$, and (b) the survival probability from the initial plane waves $2p_{}^{\rm pw}(\tau)-1$, in the PLRB model at criticality, $a=1$.
Results are shown for several values of $b<1$ at strong multifractality, and different system sizes $L$, as indicated in the legend [the same parameters as in Figs.~\ref{fig_7}(c)--\ref{fig_7}(e)].
All other features are identical to those in Fig.~\ref{fig_8}.
}\label{fig_9}
\end{figure*}

\begin{figure*}[!b]
\centering
\includegraphics[width=2\columnwidth]{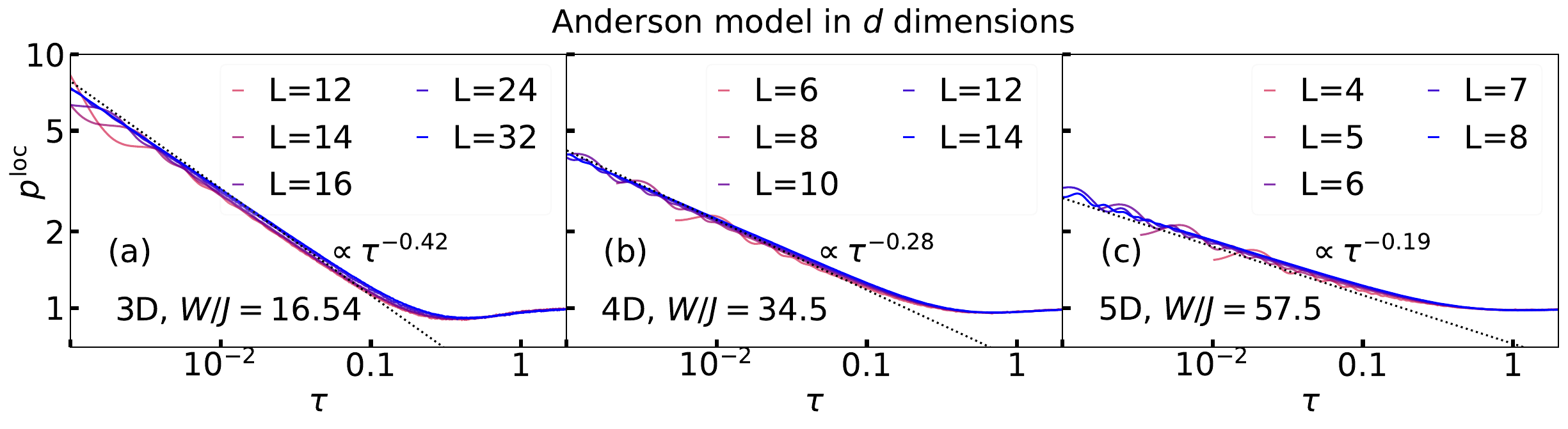}
\caption{
Survival probability $p_{}^{\rm loc}(\tau)$ from Eq.~\eqref{eq:sur_prob_norm} in the Anderson models at the critical points, for different linear system sizes $L$.
Different panels correspond to different dimensionalities $d$ of the hypercubic lattice:
(a) $d=3$ (the critical point $W/J=16.54$), (b) $d=4$ (the critical point $W/J=34.5$),  and (c) $d=5$ (the critical point $W/J=57.5$).
The dotted lines are the fits $\propto \tau^{-\beta}$ from Eq.$~$(\ref{def_taubeta}), and in each panel we provide the value of $\beta$ obtained from the fit.
}\label{fig_10}
\end{figure*}

As discussed in Sec.~\ref{sec:arguments}, the survival probability $p_{}^{\rm loc}(\tau)$ from Eq.~\eqref{eq:sur_prob_norm} is expected to exhibit a scale-invariant behavior at criticality, starting from the initial site-localized states.
In particular, $p_{}^{\rm loc}(\tau)$ is expected to follow a power-law decay from Eq.~(\ref{def_taubeta}) at mid-times, which is connected to the fractal dimension $\gamma$ via Eq.~(\ref{eq:def_gamma_beta}).

While Ref.~\cite{hopjan2023} demonstrated these properties for the 3D Anderson model, we complement them here by the study of the PLRB model.
Figure~\ref{fig_7} shows results for $p_{}^{\rm loc}(\tau)$ at the critical point $a=1$ for different system sizes $L$ and different values of the parameter $b$.
Results confirm the two expectations raised above:
(a) emergence of scale-invariant dynamics at both mid-times and late-times, and
(b) a power-law decay $p_{}^{\rm loc}(\tau) \propto \tau^{-\beta}$ in the mid-time dynamics, described by Eq.~(\ref{def_taubeta}).

\begin{figure*}[!t]
\centering
\includegraphics[width=2\columnwidth]{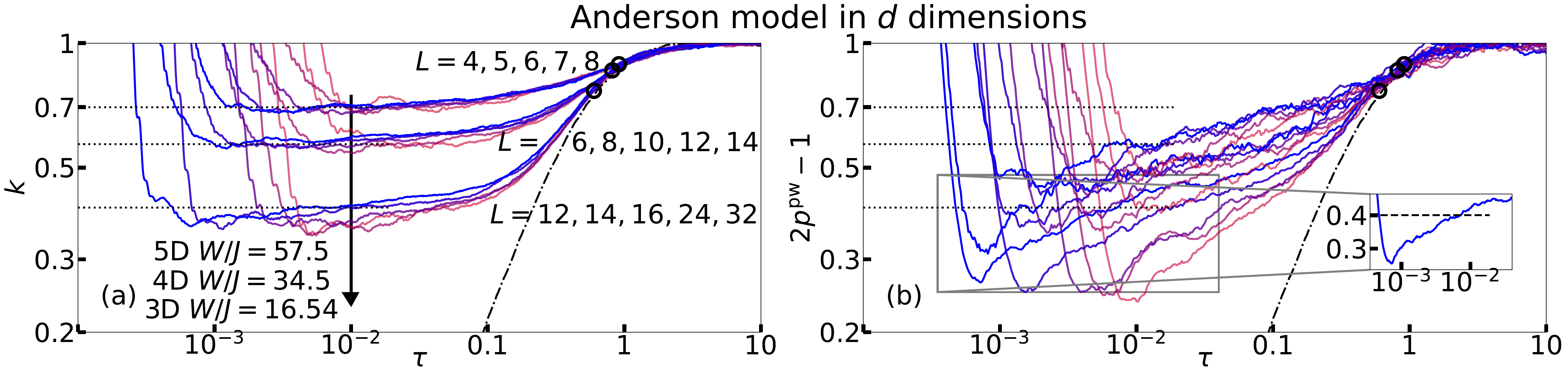}
\caption{
(a) The SFF $k(\tau)$, and (b) the survival probability from the initial plane waves $2p_{}^{\rm pw}(\tau)-1$, in the Anderson models at the critical points, for different linear system sizes $L$.
Results are shown for different dimensionalities $d$ of the hypercubic lattice: $d=3$ (the critical point $W/J=16.54$), $d=4$ (the critical point $W/J=34.5$), and $d=5$ (the critical point $W/J=57.5$).
The horizontal dotted lines denote the plateau values $k_{\chi}^{}$, and the dashed-dotted lines denote the linear ramp of the numerically evaluated GOE Hamiltonians, cf.~Fig.$~$\ref{fig_3}.
The circles correspond to the extracted Thouless times $\tau_{\rm Th}$, at which $k^{}(\tau)$ approaches the numerically evaluated GOE values. 
The inset in (b) shows a detail of the additional short-time feature of $2p_{}^{\rm pw}(\tau)-1$ compared to $k^{}(\tau)$, which approaches a plateau.
}\label{fig_11}
\end{figure*}

Different panels of Fig.~\ref{fig_7} correspond to different values of the parameter $b$, and in each panel we list the value of the power-law exponent $\beta$ obtained from the fit.
The later is connected to the fractal dimension $\gamma$ via $\gamma=n\beta$~(\ref{eq:def_gamma_beta}), where $n\approx1$ for the PLRB model.
We observe that $\beta$, and hence $\gamma$, increases with increasing $b$, i.e., weaker multifractality is associated with larger $b$.
The dependence of $\beta$ on $b$ is summarized in Fig.$~$\ref{fig_12}(a) (see below).

We next study survival probabilities from other initial states, specifically, the SFF $k^{}(\tau)$ and the survival probability from initial plane waves $p^{\rm pw}(\tau)$.
In Fig.$~$\ref{fig_4}(c) we studied the corresponding unscaled quantities $K(\tau)$ and $P^{\rm pw}(\tau)$ at criticality, which were shown to approach at late times ($\tau\to1$) the values $K\to 1/D$ and $P_{}^{\rm pw} \to 2/D$.
Their values match the predictions of the GOE Hamiltonians studied in Sec.~\ref{sec:GOE}.
For the quantities $k^{}(\tau)$ and $p^{\rm pw}(\tau)$, which are of interest here, the GOE case implies the relationship $k^{}=2p_{}^{\rm pw}-1$, as established in Eq.~\eqref{eq:K_p_relation}.
Therefore, in Figs.~\ref{fig_8} and~\ref{fig_9} we explore to which degree the similarity between $k^{}(\tau)$ and $2p_{}^{\rm pw}(\tau)-1$ extends to the dynamics at criticality.

Figure~\ref{fig_8} shows the results at weak multifractality, $b>1$, and Fig.~\ref{fig_9} shows the results at strong multifractality, $b<1$.
We observe emergence of a broad plateau in $k(\tau)$, which is marked by the horizontal dotted lines in Figs.~\ref{fig_8}(a) and~\ref{fig_9}(a).
The plateau emerges in the mid-time dynamics, in which $p_{}^{\rm loc}(\tau)$, see Fig.~\ref{fig_7}, exhibits a power-law decay.
This was first observed in~\cite{hopjan2023} for the 3D Anderson model and is here complemented by the analysis in the PLRB model.

We extract the two features of $k(\tau)$ from Figs.~\ref{fig_8}(a) and~\ref{fig_9}(a), which we  then further analyze in Fig.~\ref{fig_12} of Sec.~\ref{sec:mid_vs_late}:
the value of $k^{}$ at the plateau $k_\chi$, where $\chi$ refers to spectral compressibility$~$\cite{Evers2000, Mirlin00} responsible for the plateau, and the value of $k^{}$ at the Thouless time $k^{}(\tau_{Th})$. 
The former characterizes the mid-time dynamics, see also the discussion in Sec.~\ref{sec:mid_vs_late}, while the latter characterizes the late-time dynamics.
Both values are $b$-dependent, and they decrease with increasing $b$.

It is interesting to observe that $2p_{}^{\rm pw}(\tau)-1$ at criticality indeed shares many similarities with $k(\tau)$. 
This similarity is in particular apparent at weak multifractality, see Fig.~\ref{fig_8}, for which it persists both in the mid-time and late-time dynamics.
This similarity can be interpreted as being a consequence of proximity of the weak multifractal regime to the chaotic behavior.
Still, it contains several nontrivial implications, e.g., the extraction of Thouless time can be reliably carried out from $2p_{}^{\rm pw}(\tau)-1$.
At strong multifractality, $2p_{}^{\rm pw}(\tau)-1$ exhibits additional features in the mid-time dynamics, as shown in the insets in Figs.~\ref{fig_8}(b) and~\ref{fig_9}(b).
In particular, we observe a dip below the plateau value $k_\chi$, which is also associated with the breakdown of scale-invariant behavior.

\subsection{Critical dynamics in the Anderson models}

We complement results for the PLRB model with those for the Anderson models in hypercubes with dimensions $d=3,4,5$.
Figure~\ref{fig_10} shows the dynamics of the survival probability from the initial site-localized states $p_{}^{\rm loc}(\tau)$.
While the results in $d=3$ were already reported in Ref.~\cite{hopjan2023}, the results in $d=4$ and 5, see Figs.~\ref{fig_10}(b) and~\ref{fig_10}(c), respectively, establish generality of scale-invariant mid-time and late-time dynamics at criticality.
The exponent $\beta$ of the power-law decay in the mid-time dynamics, $p_{}^{\rm loc}(\tau) \propto \tau^{-\beta}$, decreases with increasing the dimensionality $d$, which implies stronger multifractality.
The shift towards stronger multifractality in higher-dimensional Anderson models was previously studied from the perspective of level statistics~\cite{Tarquini17} and optical conductivity~\cite{Herbrych21}.

In Fig.~\ref{fig_11} we then study the survival probabilities from other initial states, specifically, the SFF $k(\tau)$ and the survival probability from the initial plane waves $p_{}^{\rm pw}(\tau)$.
The SFF $k(\tau)$, see Fig.~\ref{fig_11}(a), exhibits a plateau in the mid-time dynamics, which is consistent with the results for the PLRB model in Figs.~\ref{fig_8}(a) and~\ref{fig_9}(a), and with the results for the 3D Anderson model in Ref.~\cite{suntajs_prosen_21}.
We note, however, that scale invariance of the plateau in Fig.~\ref{fig_11}(a) is reasonably good but not perfect, which we attribute to the absence of spectral unfolding and filtering in the calculation of the SFF $k(\tau)$.
In Fig.~\ref{fig_21} of Appendix~\ref{app:1} we show results for the SFF after spectral unfolding and filtering, for which scale invariance of the plateau is further improved.
In Fig.$~$\ref{fig_12} (see below) we summarize the behavior of $\beta$, $k_{\chi}$ in the mid-time dynamics, and $\overline{r}$, $k(\tau_{Th})$ in the late-time dynamics, for the 3D, 4D, and 5D Anderson models.

Finally, motivated by the observation of similarities between $k(\tau)$ and $2p^{\rm loc}(\tau)-1$ in the PLRB model in Figs.~\ref{fig_8} and~\ref{fig_9}, we ask to which extent this similarity translates to the Anderson models.
Results for the latter are shown in Fig.~\ref{fig_11}(b).
They share certain similarities with the PLRB model at strong multifractality, see Fig.~\ref{fig_9}(b).
Nevertheless, they do not show convincing evidence neither of scale invariance nor of building a plateau in the mid-time dynamics, at least for the system sizes of investigation.
Still, based on the similarity with the PLRB models, for which the latter properties appear to be present at weak multifractality, see Fig.~\ref{fig_8}(b), we can not exclude that they also appear at strong multifractality in system sizes much larger than those considered here.

\begin{figure}[!b]
\centering
\includegraphics[width=\columnwidth]{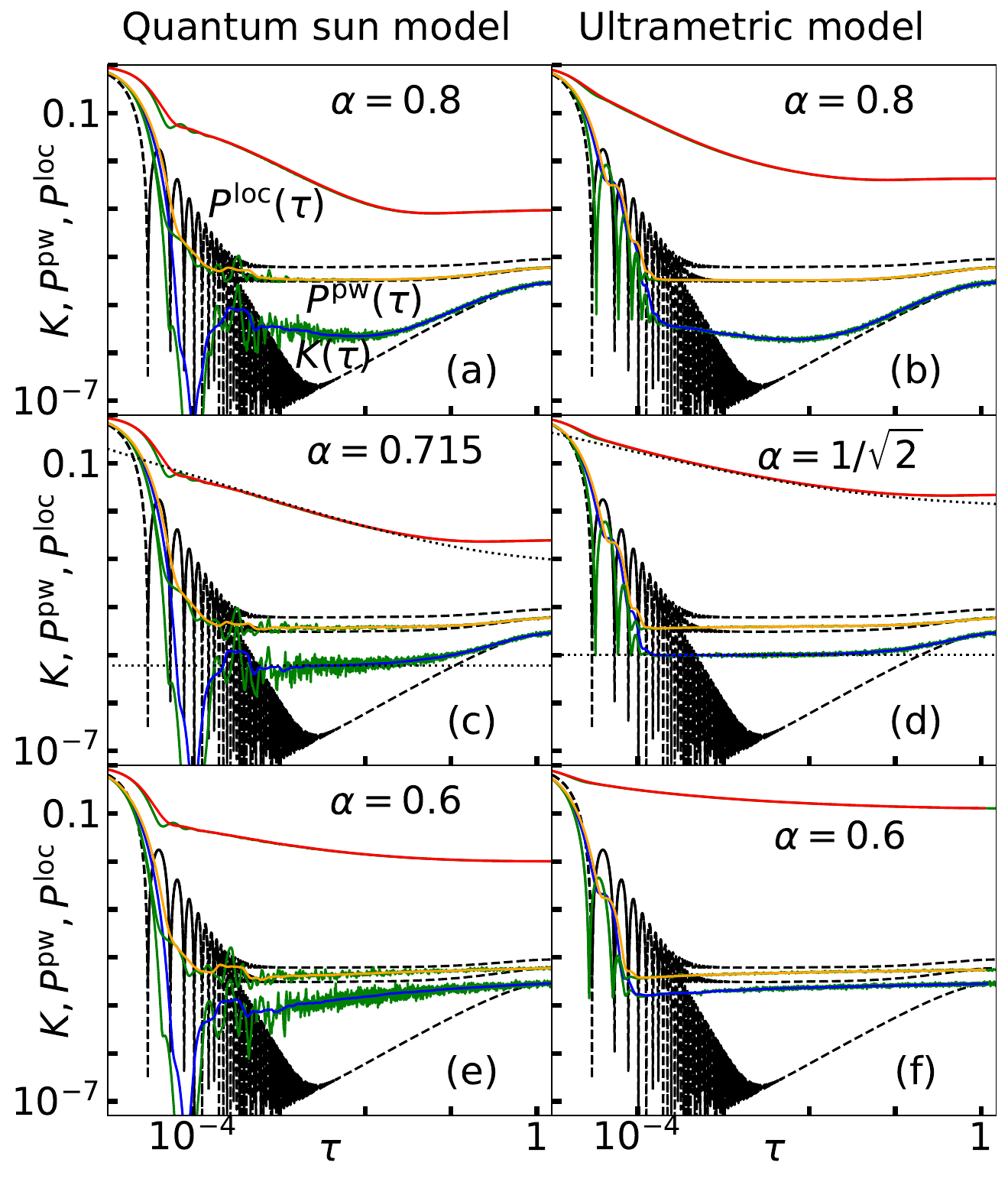}
\caption{
$K(\tau)$, $P^{\rm loc}(\tau)$, and $P^{\rm pw}(\tau)$ (all green) in the QS model (left column) and UM model (right column) at $L+N=15$, i.e., $D=2^{15}=32768$.
[(a), (b)] Ergodic phase at $\alpha=0.8$, [(c), (d)] critical point at $\alpha_c$ [$\alpha_c = 0.715$ in the QS model at $N=5$~\cite{hopjan2023} and $\alpha_c=1/\sqrt{2}$ in the UM model], [(e), (f)] localized phase at $\alpha=0.6$.
Additionally, we show the running averages to $K(\tau)$ [blue], $P_{}^{\rm loc}(\tau)$ [red] and $P_{}^{\rm pw}(\tau)$ [orange]. 
The corresponding results for the GOE Hamiltonians from Sec.~\ref{sec:GOE} (see Fig.~\ref{fig_2}) are shown as dashed black lines. 
The dotted lines in (c) and (d) indicate the power-law decay for $P_{}^{\rm loc}(\tau)$ to a nonzero constant and the plateaus in $K(\tau)$.
}\label{fig_13}
\end{figure}

\begin{figure*}[!t]
\centering
\includegraphics[width=2\columnwidth]{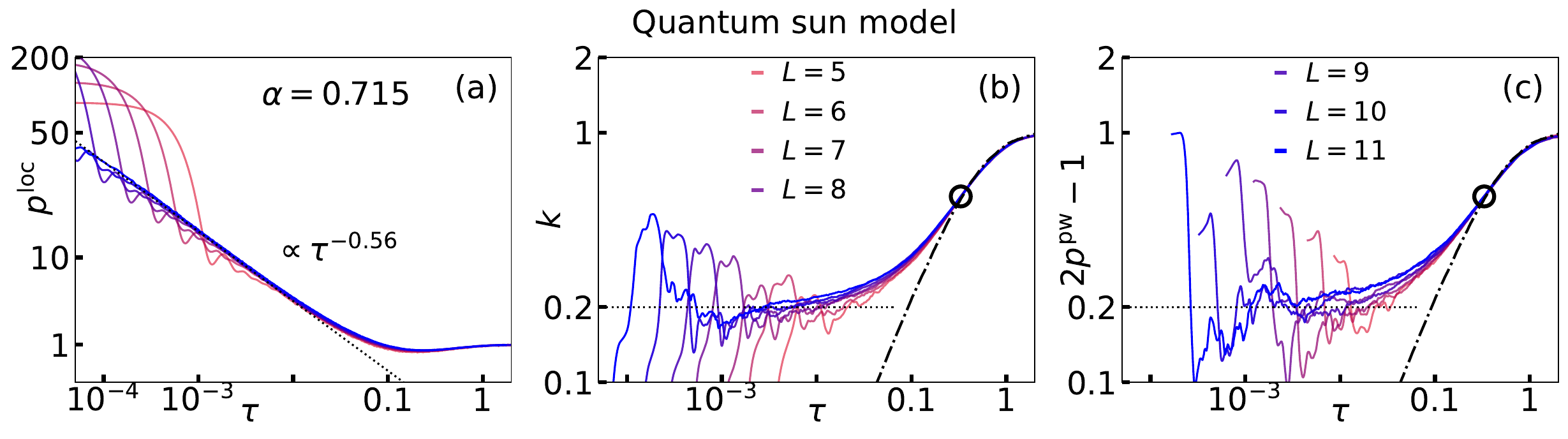}
\caption{
Survival probabilities in the QS model at criticality, $\alpha=\alpha_c = 0.715$, for several system sizes $L$ with the Fock-space dimension $D=2^{N+L}$ and $N=5$.
(a) $p_{}^{\rm loc}(\tau)$ from the initially Fock-space-localized states,
(b) the SFF $k^{}(\tau)$, and (c) $2p_{}^{\rm pw}-1$ from the initial plane waves.
The dotted line in (a) denotes the power-law fit from Eq.$~$(\ref{def_taubeta}) and the horizontal dotted lines in (b) and (c) denote the plateau values $k_{\chi}^{}$. 
The thick dashed-dotted lines in (b) and (c) denote the numerical results for the SFF of the GOE Hamiltonians, cf.~Fig.$~$\ref{fig_3}. 
The black circles correspond to the extracted Thouless times $\tau_{\rm Th}$.
}\label{fig_14}
\end{figure*}

\begin{figure*}[!t]
\centering
\includegraphics[width=2\columnwidth]{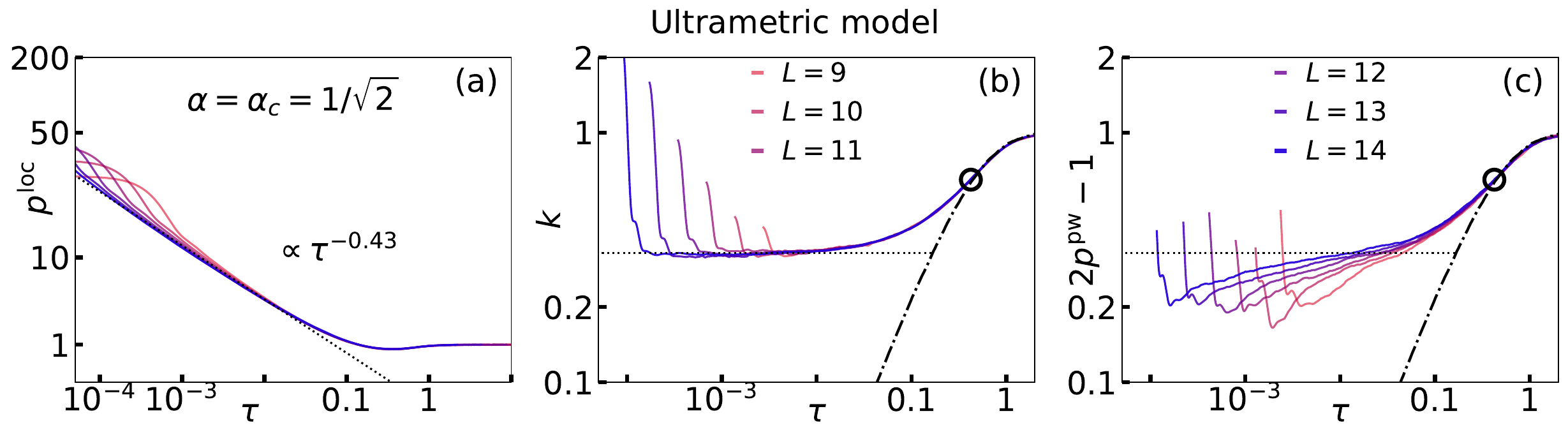}
\caption{
Survival probabilities in the UM model at criticality, $\alpha=\alpha_c = 1/\sqrt{2}$, for several system sizes $L$ with the Fock-space dimension $D=2^{N+L}$ and $N=1$.
(a) $p_{}^{\rm loc}(\tau)$ from the initially Fock-space-localized states,
(b) the SFF $k^{}(\tau)$, and (c) $2p_{}^{\rm pw}-1$ from the initial plane waves.
The dotted line in (a) denotes the power-law fit from Eq.$~$(\ref{def_taubeta}) and the horizontal dotted lines in (b) and (c) denote the plateau values $k_{\chi}^{}$. 
The thick dashed-dotted lines in (b) and (c) denote the numerical results for the SFF of the GOE Hamiltonians, cf.~Fig.$~$\ref{fig_3}. 
The black circles correspond to the extracted Thouless times $\tau_{\rm Th}$.
}\label{fig_15}
\end{figure*}

\section{Results for interacting models} \label{sec:Avalanche}

\subsection{From many-body quantum chaos to Fock space localization} \label{sec:QS_UM}

We now study the dynamics of two interacting models, the QS model~(\ref{eq:hamQSM}) and the UM model~(\ref{eq:def_rmt_model}).
The initial states are many-body states, which belong to the Fock space of dimension $D=2^{L+N}$.
Previous study has established that both models exhibit an ergodicity breaking phase transition from many-body quantum chaos at $\alpha>\alpha_c$ to Fock space localization at $\alpha<\alpha_c$, with the critical point exhibiting multifractality~\cite{suntajs2023similarity}.

We first study the dynamics across the critical point at a fixed system size $L$, focusing on the unscaled SFF $K(\tau)$ from Eq.~(\ref{eq:sff_raw}) and the survival probability $P(\tau)$ from Eq.~(\ref{eq:sur_averaged}).
Figure~\ref{fig_13} compares the dynamics in both models.
The results exhibit strong similarities in all phases, in particular in the mid-time and late-time dynamics.
These similarities can be interpreted as the extension of the established similarity between the two models in Ref.~\cite{suntajs2023similarity} to time domain.
In the ergodic phase, see Figs.~\ref{fig_13}(a) and~\ref{fig_13}(b), $K(\tau)$ and $P^{\rm pw}(\tau)$ are close to the GOE predictions, while $P^{\rm loc}(\tau)$ is not.
At criticality, see Figs.~\ref{fig_13}(c) and~\ref{fig_13}(d), $P^{\rm loc}(\tau)$ develops a power-law decay and $K(\tau)$ exhibits a plateau.
In the localized phase, see Figs.~\ref{fig_13}(e) and~\ref{fig_13}(f), properties of the critical point gradually fade away in both models.

\subsection{Similarity of critical dynamics in QS and UM models} \label{sec:QS_UM_similarity}

We next focus our analysis on the dynamics at criticality, $\alpha=\alpha_c$.
As before, we study survival probabilities introduced in Eq.~(\ref{eq:sur_prob_norm}) from different initial states, $p^{\rm loc}(\tau)$, $k(\tau)$, and $p^{\rm pw}(\tau)$.

The survival probability from the initial Fock-space-localized states $p^{\rm loc}(\tau)$ exhibits a scale-invariant power-law decay in the mid-time dynamics of both models, shown in Figs.~\ref{fig_14}(a) and~\ref{fig_15}(a).
The exponent $\beta$ of the power-law decay, however, depends on the specific properties of the models.
For the model parameters set in Sec.~\ref{sec:models}, the systems are at moderate to strong multifractality. 

The SFF $k(\tau)$ exhibits a scale-invariant plateau in both models, see Figs.~\ref{fig_14}(b) and~\ref{fig_15}(b).
Scale invariance of the plateau is well converged in the UM model in Fig.~\ref{fig_15}(b).
On the other hand, in the QS model the tendency towards scale invariance is apparent but the convergence is not yet optimal.
In Ref.~\cite{suntajs_vidmar_22} the SFF was studied using spectral unfolding and filtering, giving rise to clearer convergence towards scale-invariant behavior.

For the survival probability from the initial plane waves $p^{\rm pw}(\tau)$ we study its rescaled version $2p^{\rm pw}(\tau)-1$, which is identical to the SFF $k(\tau)$ in the GOE Hamiltonians~(\ref{eq:K_p_relation}).
This relationship also holds for both models at criticality in the late-time dynamics, see Figs.~\ref{fig_14}(c) and~\ref{fig_15}(c).
As a consequence, the Thouless times, which are very similar in both models, can be extracted either from $k(\tau)$ or from $2p^{\rm pw}(\tau)-1$.
However, at mid-times the validity of the relationship between $k(\tau)$ and $2p^{\rm pw}(\tau)-1$ is less obvious due to the emergence of dips in $2p^{\rm pw}(\tau)-1$ below the plateau values.

\subsection{Similarity with quadratic models}

Apart from the similarity in the mid-time and late-time dynamics in the interacting QS and UM models, there is also an apparent similarity between the interacting models considered here and the quadratic models (the PLRB and Anderson models) studied in Sec.~\ref{sec:Anderson_PLRB}.
The emergence of this similarity is nontrivial provided that in the latter case one considers single-particle states and single-particle spectrum, while in the former case all quantities are of genuinely many-body origin.

The most prominent features that exhibit similarity in the mid-time dynamics are the scale-invariant power-law decay of $p^{\rm loc}(\tau)\propto \tau^{-\beta}$ and the plateau value $k_{\chi}$ of the SFF $k(\tau)$.
We characterize them by studying the behavior of $\beta$ and $1-k_{\chi}$ in Fig.~\ref{fig_12}(a)  (see below) as a function of the degree of multifractality (see Sec.~\ref{sec:mid_vs_late} for details).
The results for the interacting and quadratic models exhibit, for each of the measures $\beta$ and $1-k_{\chi}$, a nearly perfect collapse to a single function.

While the above properties hint at the universality in mid-time dynamics, it is interesting to note that quantitatively similar behavior can also be observed in the late-time dynamics.
In particular, in Fig.~\ref{fig_12}(b) (see below) we plot $1-k(\tau_{Th})$, i.e., the distance of the SFF $k(\tau)$ at the onset of quantum chaos to its long-time average, and again observe a nearly perfect collapse.
This suggests that fingerprints of criticality may be encoded in both the mid-time as well as the late-time dynamics.

\begin{figure}[t!]
\centering
\includegraphics[width=0.90\columnwidth]{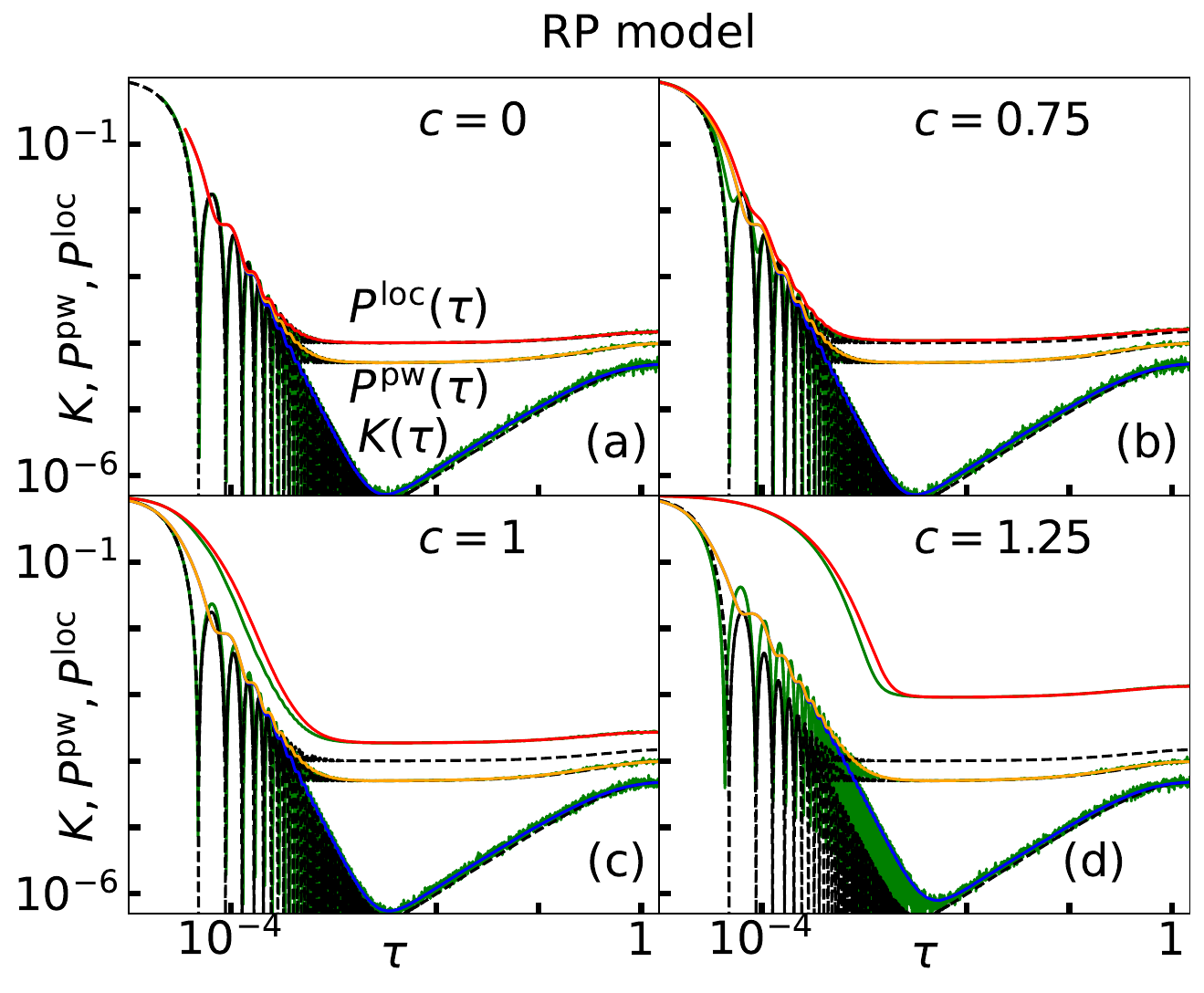}
\caption{
$K(\tau)$, $P^{\rm loc}(\tau)$, and $P^{\rm pw}(\tau)$ (all green) in the RP model at $\lambda=0.5$ and $D=L=20000$.
(a) $c=0$, (b) $c=0.75$, (c) the first critical point $c=1$, and (d) $c=1.25$. Additionally, we show the running averages to $K$ (blue), $P_{}^{\rm loc}$ (red), and $P_{}^{\rm pw}$ (orange). 
The corresponding results for the GOE Hamiltonians from Sec.~\ref{sec:GOE} (see Fig.~\ref{fig_2}) are shown as dashed black lines. 
At $c=1$ the SFF $K^{}(\tau)$ fully matches the GOE prediction.
}\label{fig_16}
\end{figure}

\begin{figure}[b!]
\centering
\includegraphics[width=0.90\columnwidth]{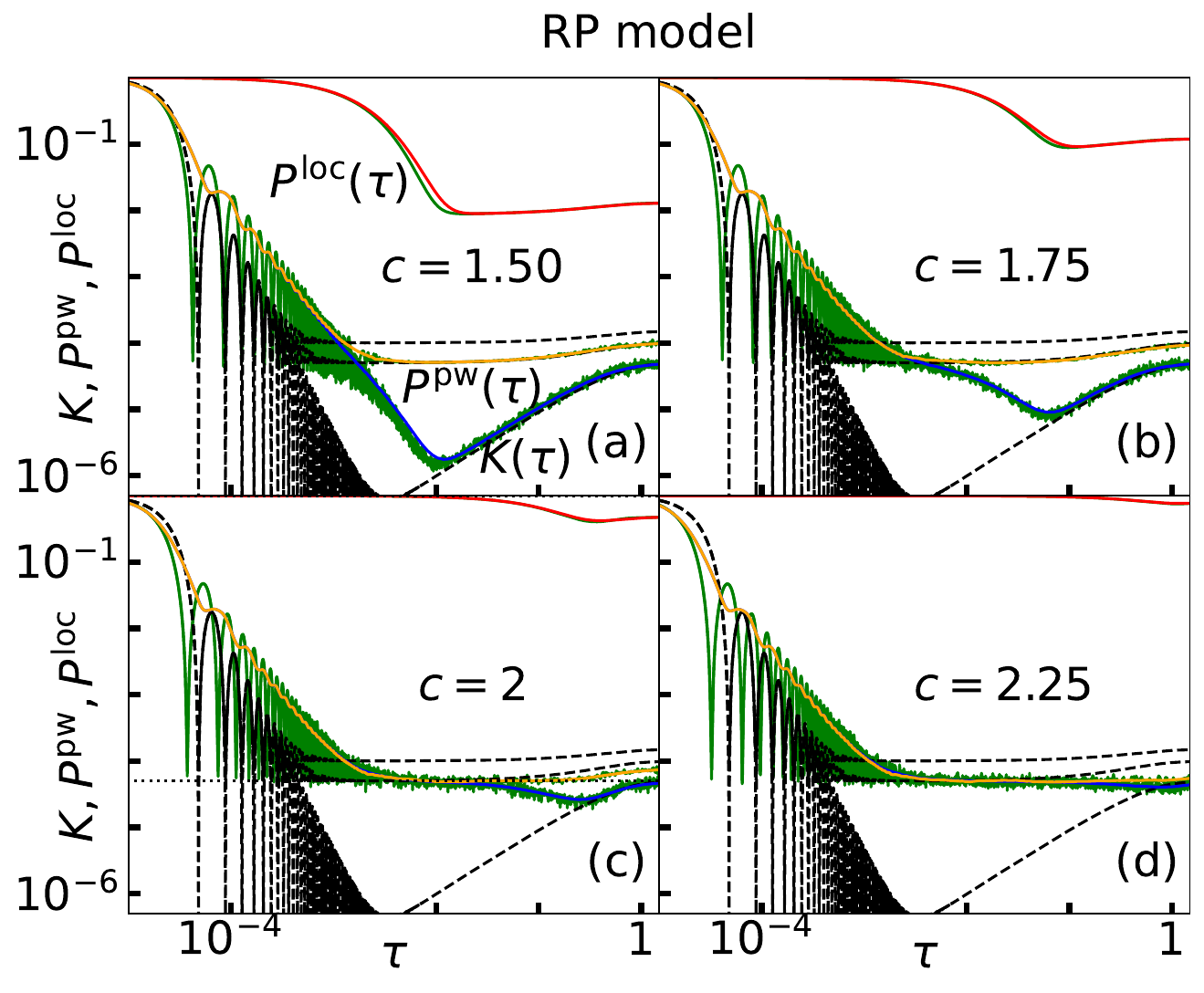}
\caption{
$K(\tau)$, $P^{\rm loc}(\tau)$, and $P^{\rm pw}(\tau)$ (all green) in the RP model at $\lambda=0.5$ and $D=L=20000$.
(a) $c=1.5$, (b) $c=1.75$, (c) the second critical point $c=2$, and (d) $c=2.25$. Additionally, we show the running averages to $K$ (blue), $P_{}^{\rm loc}$ (red), and $P_{}^{\rm pw}$ (orange). 
The corresponding results for the GOE Hamiltonians from Sec.~\ref{sec:GOE} (see Fig.~\ref{fig_2}) are shown as dashed black lines. 
The dotted lines in (c) indicate the plateau in $K(\tau)=1/D$.
}\label{fig_17}
\end{figure}

\section{Results for the Rosenzweig-Porter (RP) model} \label{sec:RPmodel}

\begin{figure*}[!t]
\centering
\includegraphics[width=2\columnwidth]{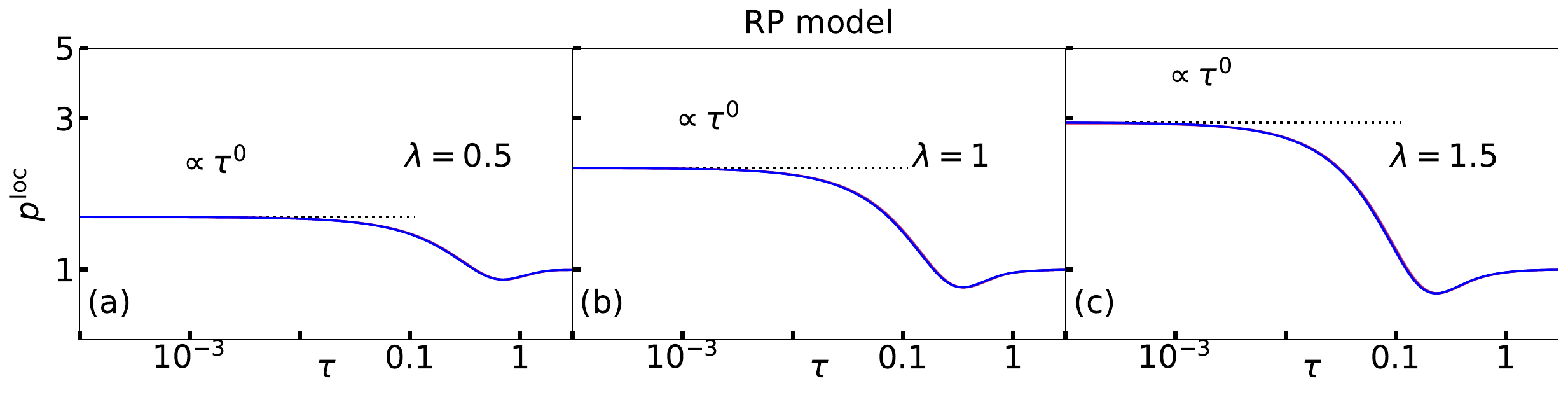}
\caption{
Survival probability $p_{}^{\rm loc}(\tau)$ from Eq.~\eqref{eq:sur_prob_norm_rp} in the RP model at the critical point $c=2$.
(a) $\lambda=0.5$, (b) $\lambda=1$, and (c) $\lambda=1.5$.
The horizontal dotted lines can be interpreted as the fits $p_{}^{\rm loc}(\tau)\propto \tau^{-\beta}$ from Eq.$~$(\ref{def_taubeta}), with $\beta=0$ in all panels.
Results are shown for different linear system sizes $L=$ 1000, 2500, 5000, 10 000, 20 000, which perfectly overlap.
}\label{fig_18}
\end{figure*}

\begin{figure*}[!t]
\centering
\includegraphics[width=2\columnwidth]{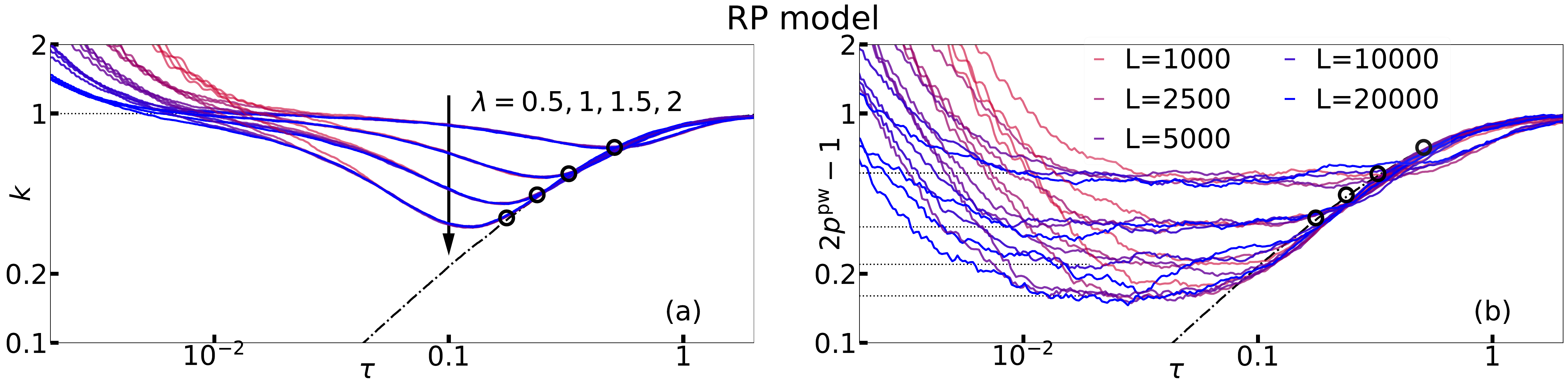}
\caption{
(a) The SFF $k(\tau)$, and (b) the survival probability from the initial plane waves $2p_{}^{\rm pw}(\tau)-1$, in the RP model at the critical point $c=2$, for different linear system sizes $L$.
Results are shown for different values of $\lambda$, as indicated in the legend.
The horizontal dotted line in (a) denotes the plateau value $k_{\chi}^{}=1$, and the horizontal dotted lines in (b) denote the estimates of the plateau values for $2p_{}^{\rm pw}(\tau)-1$. 
The dashed-dotted lines in both panels denote the linear ramp of the numerically evaluated GOE matrices, cf.~Fig.$~$\ref{fig_3}.
The circles correspond to the extracted Thouless times $\tau_{\rm Th}$, at which the quantities approach the numerically evaluated GOE value. 
}\label{fig_19}
\end{figure*}

Previous two sections have established similarity in the critical dynamics in the quadratic and interacting models.
Here we show that the RP model exhibits a slightly different behavior that, although consistent internally, features distinct mid-time critical dynamics at the eigenstate transition.

\subsection{Transitions in the RP model} \label{sec:RP_trans}

Even though we defined the RP model in Sec.~\ref{sec:def_RPmodel} as a quadratic model with random on-site potentials and all-to-all hoppings, the model is different from the other models considered so far since it is expected to exhibit two transitions, at $c=1$ and $c=2$.
In the context of this paper we argue that only the second transition (at $c=2$) can be considered as the eigenstate transition with scale-invariant critical dynamics.
At this transition, the short-range spectral correlations change from the GOE statistics to the Poisson statistics at any $\lambda$~\cite{Pino19}. 
Here we are mostly concerned with the properties of long-range spectral correlations~\cite{Kravtsov15, DeTomasi19, Barney23,Buijsman23}.

We first study the dynamics at and in the vicinity of the first transition at $c=1$.
In Fig.$~$\ref{fig_16}, we show an example of the unscaled survival probability $P(\tau)$ from Eq.$~$\eqref{eq:sur_averaged} as well as the SFF $K(\tau)$.
We vary the parameter $c$ while keeping the parameter $\lambda$ fixed.
Figure~\ref{fig_16}(a) shows that the limit of $c=0$ is, up to normalization, the limit of the GOE Hamiltonians, and the numerical simulations fit the corresponding analytical predictions very well.
Increasing the value of $c$ within the range $c\leq 1$, see Figs.~\ref{fig_16}(b) and~\ref{fig_16}(c), gives rise to the visible differences in $P^{\rm loc}(\tau)$, while $K(\tau)$ and $P^{\rm pw}(\tau)$ remain virtually identical.
Above the transition point to the non-ergodic extended phase, i.e., at $c>1$ shown in Fig.$~$\ref{fig_4}(d), the deviations of $K(\tau)$ and $P^{\rm pw}(\tau)$ from the analytical results become more apparent.

Next we study the dynamics at and in the vicinity of the second transition at $c=2$.
By increasing $c$ toward the value $c=2$, the results for $K(\tau)$ and $P^{\rm pw}(\tau)$ further deviate from the GOE predictions, see Figs.~\ref{fig_17}(a)--\ref{fig_17}(c).
In particular, the agreement between $K(\tau)$ and the GOE predictions is only preserved at long times close to the Heisenberg time $\tau_H = 1$.
It is interesting to note that by increasing $c$, $P^{\rm pw}(\tau)$ approaches the time dependence of $K(\tau)$, and at $c>2$, see Fig.~\ref{fig_17}(d), they indeed become very similar.
This excludes the validity of the relationship $k^{}(\tau) = 2p_{}^{\rm pw}(\tau)-1$ from Eq.~(\ref{eq:K_p_relation}), which is further confirmed in the next section where we study scale invariance of the dynamics at $c=2$.

Another important aspect concerns the mid-time dynamics of $K(\tau)$.
Figure~\ref{fig_17}(c) shows that at $c=2$, which we refer to here as the eigenstate transition, $K(\tau)$ develops a plateau, similarly to the critical points for eigenstate transitions in other models under considerations. 
However, the plateau value is maximal in the sense that it equals the long-time average. 
As discussed in the next section, this is consistent with the absence of the decay of $P^{\rm loc}(\tau)$ in the mid-time dynamics, which is indeed observed in Fig.~\ref{fig_17}(c).

\begin{figure*}[!t]
\centering
\includegraphics[width=2\columnwidth]{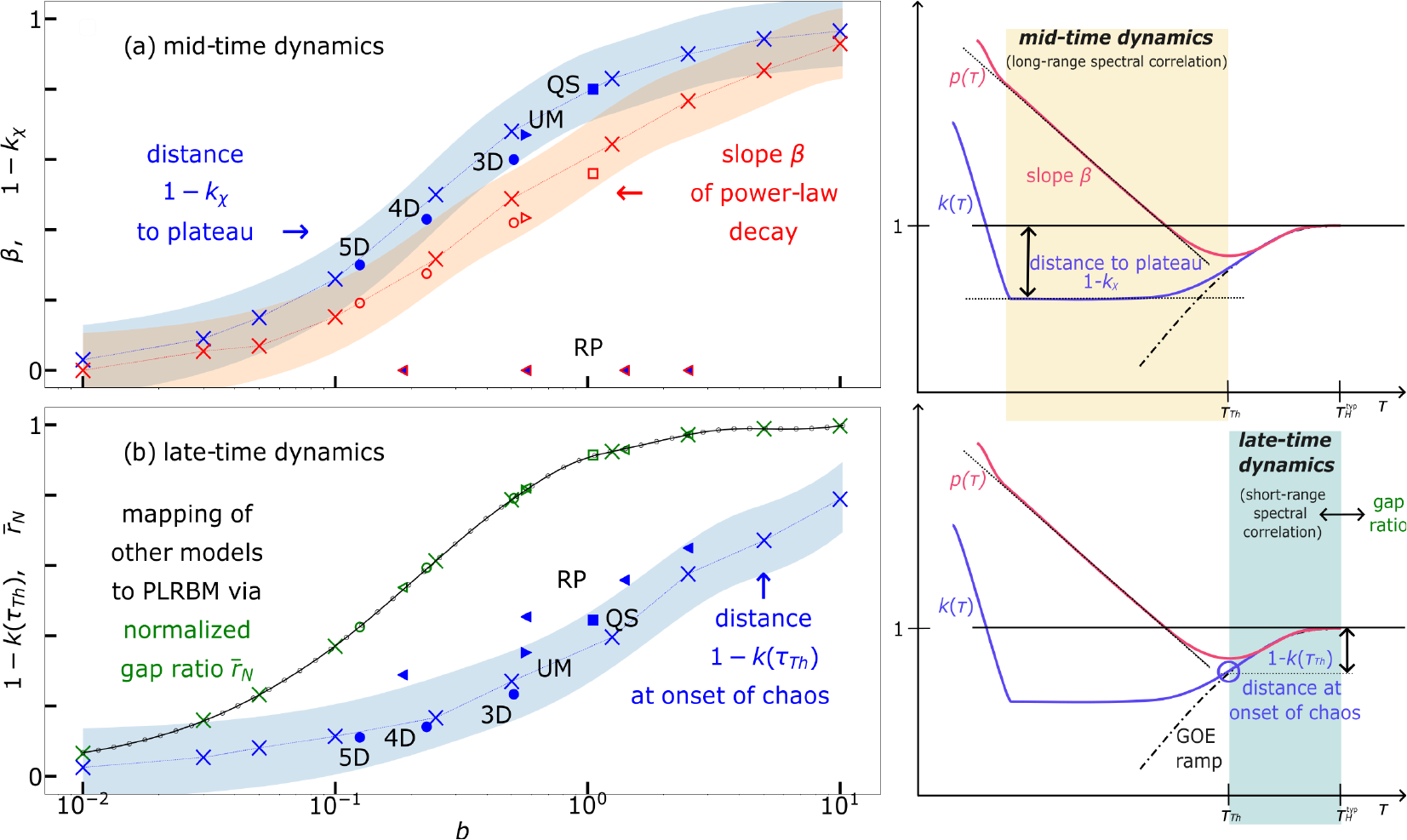}
\caption{Characterization of scale-invariant properties at criticality in the (a) mid-time dynamics, and (b) late-time dynamics, see also Fig.$~$\ref{fig_1}.
(a) The power-law decay exponent $\beta$ (red) and the distance of the SFF long-time average to the plateau, $1-k_\chi$ (blue).
(b) The distance of the SFF long-time average to the SFF at the onset of quantum chaos, $1-k(\tau_{\rm Th})$ (blue), and the normalized average gap ratio $\overline{r}_N$ from Eq.~(\ref{eq:r_normalized}) (black). 
The crosses ($\times$) are results for the PLRB model, as a function of the parameter $b$.
For all other models, we assign the corresponding value of $b$ by matching the values of $\overline{r}_N$ in the studied model with those in the PLRB model.
This yields the $\overline{r}_N$ versus $b$ curve smooth by construction.
Results for the 3D, 4D and 5D Anderson models are marked with circles ($\circ$), for the QS model with squares ($\square$), for the UM model with right triangles ($\triangleright$), and for the RP model with left triangles ($\triangleleft$).
}\label{fig_12}
\end{figure*}

\subsection{Critical dynamics in the RP model} \label{sec:RP}

We now focus on the dynamics at the critical point $c=2$, and we study the survival probabilities $p^{\rm loc}(\tau)$ in Fig.~\ref{fig_18}, and $k(\tau)$, $p^{\rm pw}(\tau)$ in Fig.~\ref{fig_19}.
The most important property of the critical point at $c=2$ is that Hamiltonian eigenstates are localized~\cite{Kravtsov15, DeTomasi19}, in contrast to all other critical points considered before.
This has implications both for the time dependence in the mid-time dynamics, as well as on the scale invariance since it requires a refinement of the definition of the survival probability $p^{\rm loc}(\tau)$.

As a consequence of localization at the critical point one needs to consider Eq.~(\ref{def_Pbar}) with some care.
For the initial site-localized states, we observe $\overline{P_{}^{\rm loc}} = P_\infty^{\rm loc}$ for any finite system under consideration.
Then, there is no need to apply Eq.~\eqref{eq:sur_prob_norm}, but a simpler rescaling,
\begin{equation}
\label{eq:sur_prob_norm_rp}
p^{\rm loc}(\tau)= \frac{P^{\rm loc}(\tau)}{\overline{P^{\rm loc}}} \;,
\end{equation}
which is analogous to the one applied to the SFF in Eq.~(\ref{def_ktau}), is sufficient to observe scale invariance.
The resulting $p^{\rm loc}(\tau)$ is plotted in Fig.$~$\ref{fig_18} at $c=2$ and several values of $\lambda$, exhibiting perfect scale invariance.

Another important feature of Fig.$~$\ref{fig_18} is the absence of any decay of $p^{\rm loc}(\tau)$ in the mid-time dynamics for all values of $\lambda$.
This can be interpreted as a power-law decay with $\beta=0$, and hence the fractal dimension is $\gamma=0$, which is consistent with localization.
A manifestation of this property is the emergence of a plateau in $k(\tau)$ in the mid-time dynamics, shown in Fig.$~$\ref{fig_19}(a), which equals $k_\chi = 1$. We note that the plateau is more clearly visible in the unfolded and filtered version of the SFF $\tilde{k}(\tau)$ shown in Fig.$~$\ref{fig_22} of Appendix~\ref{app:1}.
A certain tendency towards scale invariance is also manifested in the dynamics of $2p_{}^{\rm pw}(\tau)-1$ shown in Fig.$~$\ref{fig_19}(b), however, as already noted in the context of Fig.~\ref{fig_17}(c), the time evolution of the later quantity can not be related to $k(\tau)$.

The main properties of the RP model are summarized in the next section, see Fig.$~$\ref{fig_12}, and compared to other models.
For the long-time dynamics, one can show that the RP model complies with the universal features observed in other models, see the next section for details.
However, the mid-time dynamics of the RP model is not consistent with the universal properties of other models.
We attribute localization of the Hamiltonian eigenfunctions as the main source of deviations, since all the other models exhibit multifractality at criticality.

\section{Mid-time vs late-time dynamics} \label{sec:mid_vs_late}

We summarize our findings by combining the two perspectives on the properties of the critical points, one obtained from the mid-time dynamics and another from the late-time dynamics.
The mid-time dynamics is characterized by the power-law exponent $\beta$ and the distance of the SFF long-time average to the plateau, $1-k_\chi$,
while the long-time dynamics is characterized by the normalized average gap ratio at criticality,
\begin{equation} \label{eq:r_normalized}
\overline{r}_N= \frac{\overline{r}-r_{P}}{r_{\rm GOE}-r_{P}} \;,
\end{equation}
and the distance of the SFF at the onset of quantum chaos (i.e., at $\tau = \tau_{\rm Th}$) to its long time average, $1-k(\tau_{\rm Th})$.
The goal is to study all these quantities as a function of a single parameter that determines the degree of multifractality of the critical wavefunctions.
We choose this parameter to be $b$ from the PLRB model, see Eq.~(\ref{eq:ham_ensemble}).
For the other models, we determine the corresponding value of $b$ by requesting the gap ratio $\overline{r}_N$ of that model to match the gap ratio of the PLRB model.

Results are shown in Fig.~\ref{fig_12}.
For the late-time dynamics, see Fig.~\ref{fig_12}(b), the curve $\overline{r}_N$ versus $b$ is smooth by construction.
It is interesting to observe that also the distance of the SFF long time average to the SFF at the onset of quantum chaos, $1-k(\tau_{\rm Th})$, appears to approach a well-defined function of the parameter $b$.
This observation is in particular important since we consider both quadratic and interacting models within the same framework.

For the mid-time dynamics, see Fig.$~$\ref{fig_12}(a), we observe that both the power-law exponent $\beta$ and the distance of the SFF to the plateau $1-k_\chi$ appear to be monotonously increasing functions of $b$.
Even though the quantities are extracted purely from the dynamics, the results for quadratic models are fully consistent with the findings of Refs.~\cite{Evers2000,Mirlin00}, where similar quantities (fractal dimension and spectral rigidity) were extracted from the analysis of energy eigenspectra and properties of eigenvectors.
Again, our main result is that also the critical properties of interacting models may belong to the same single-parameter function.

The exception to the observed universal properties of the mid-time and late-time dynamics is the RP model. 
Figure~\ref{fig_12}(a) shows that both $\beta$ and $1-k_\chi$ deviate from the single-parameter function, since $\beta=1-k_\chi=0$ in all cases under consideration.
In fact, since the wavefunctions at criticality are localized, they should rather correspond to the $b=0$ limit of the PLRB model.
This does not match with the properties of other models for which the average gap ratio is given by the Poisson statistics if critical point is localized.
The observed deviations only hint at a lack of the correspondence between the mid-time and late-time dynamics in the RP model. 

\section{Conclusions} \label{sec:conclusions}

In this paper we studied quantum dynamics of survival probabilities from different initial states, including those that define the SFF, and we provided answers to the two questions posed in the Introduction.
The most important statement is that there exist the notion of scale invariance in quantum dynamics at criticality.
Remarkably, the scale invariance is observed in quantities that bear direct analogies with the well-established measures of quantum chaos.
Another important statement is that scale invariance at criticality may be observed at times that are much shorter than those required to detect quantum chaotic dynamics.
These properties open possibilities to detect fingerprints of criticality at experimentally accessible time scales (in the so-called mid-time dynamics), which are much shorter than the longest time scales of finite systems, characterized within the notion of the Heisenberg time.

While universality of the chaotic dynamics such as the emergence of a ramp in the SFF is expected for all quantum chaotic models, understanding of the universal features at criticality is far from understood.
The survival probabilities introduced here and in the preceding Letter~\cite{hopjan2023} allow for establishing similarity of quantum dynamics in different models.
In particular, we established the link within two classes of models:
for quadratic models, between the Anderson model and the PLRB model, for which connections have often been explored in the past~\cite{EversMirlin2008}, and for interacting models, between the QS model and the UM model, for which connections have been noted only recently~\cite{suntajs2023similarity}.
Most intriguingly, however, we also established similarity in quantum dynamics between the quadratic and interacting models.
One of the possible outcomes of this result is that ergodicity breaking phase transitions in the interacting models under consideration may be described as Anderson transitions in the corresponding Hilbert spaces.

A common feature of all the models for which similarity in the critical dynamics has been established is the wavefunction structure, which is multifractal in the corresponding Hilbert spaces.
This motivated us to quantify features of both mid-time and late-time dynamics, and express them in terms of a single parameter that quantifies the degree of multifractality. 
Results shown in Fig.$~$\ref{fig_12} suggest that indeed a universal, single-parameter description of the main dynamical features is possible.
In particular, it allows for understanding the mid-time dynamics via the late-time dynamics, and vice versa.
We also showed that the RP model represents an exception from the latter property, and we attributed this by the lack of multifractal structure in the RP model at the critical point.

\acknowledgements
We acknowledge discussions with F. Heidrich-Meisner, S. Jiricek, P. \L yd\.{z}ba, P. Prelov\v{s}ek, and J. \v{S}untajs.
This work was supported by the Slovenian Research and Innovation Agency (ARIS), Research core fundings Grants No. P1-0044, N1-0273 and J1-50005.
We gratefully acknowledge the High
Performance Computing Research Infrastructure Eastern
Region (HCP RIVR) consortium~\cite{vega1} and
European High Performance Computing Joint Undertaking (EuroHPC JU)~\cite{vega2}  for funding
this research by providing computing resources of the
HPC system Vega at the Institute of Information sciences
~\cite{vega3}.

\appendix
\section{Unfolding and filtering in the SFF} \label{app:1}

The SFF studied in the main text is defined such that it allows for making the connection to the dynamics of survival probabilities.
However, the more common implementation of the SFF includes spectral unfolding and filtering.
Below we quantitatively compare both implementations of the SFF.

The SFF with unfolding and filtering~\cite{suntajs_bonca_20a, suntajs_prosen_21} is defined as 
\begin{equation} \label{def_Kt}
\tilde{k}(\tilde{\tau}) = \frac{1}{Z} \left\langle \left|\sum_{\nu = 1}^D \rho(\varepsilon_\nu) e^{-i 2\pi\varepsilon_\nu \tilde{\tau}}\right|^2 \right\rangle_H \, ,
\end{equation}
where $\{ \varepsilon_\nu \}$ denote the complete ordered set of the eigenvalues $\{ E_\nu \}$ of $\hat H$ after spectral unfolding, $\rho(\varepsilon_\nu)$ represents a Gaussian filter, and $\langle ... \rangle_H$ denotes the average over different realizations of the Hamiltonian $\hat H$.

The purpose of carrying out spectral unfolding is to eliminate the impact of non uniform density of states.
We unfold the spectrum using the cumulative spectral function ${\cal G}(E) = \sum_\nu \Theta(E-E_\nu)$, where $\Theta$ is the unit step function, and a polynomial fit $\bar g_3(E)$ of degree 3 to ${\cal G}(E)$.
The unfolded eigenvalues are then defined as $\varepsilon_\nu = \bar g_3(E_\nu)$.
Consequently, the local level spacing is set to unity at all energy densities, and the scaled time $\tilde{\tau}$ is measured in units of the inverse of this level spacing (times $\hbar$, which is also set to 1).
This means that the Heisenberg time in this units equals $\tilde{\tau}_H=1$.
One observes similarities between the scaled time $\tilde{\tau}$ and the scaled time $\tau$ from Eq.~(\ref{def_tau}), and for the purpose of the numerical studies in this appendix, we simply refer to the scaled time as $\tau$ in both cases.

The role of the Gaussian filter $\rho(\varepsilon_\nu)$ is to eliminate contributions from the spectral edges~\cite{suntajs_bonca_20a}.
The filtering function is defined as $\rho(\varepsilon_\nu) = \exp\{-\frac{(\varepsilon_\nu - \bar\varepsilon)^2}{2\eta\Gamma^2}\}$, where $\bar\varepsilon$ and $\Gamma^2$ are the mean energy and the variance, respectively, for a given Hamiltonian realization, and $\eta=0.5$ controls the effective fraction of eigenstates included in $\tilde{k}(\tau)$.
The normalization $Z = \langle \sum_\nu |\rho(\varepsilon_\nu)|^2 \rangle_H$ then sets $\tilde{k}(\tau\gg 1) \simeq 1$.
At the Thouless time $\tau_{\rm Th}$, $\tilde{k}(\tau)$ becomes universal in the sense that it matches the analytical prediction of the GOE from Eq.~(\ref{eq:K_goe_ramp}).

\begin{figure}[!b]
\centering
\includegraphics[width=0.98\columnwidth]{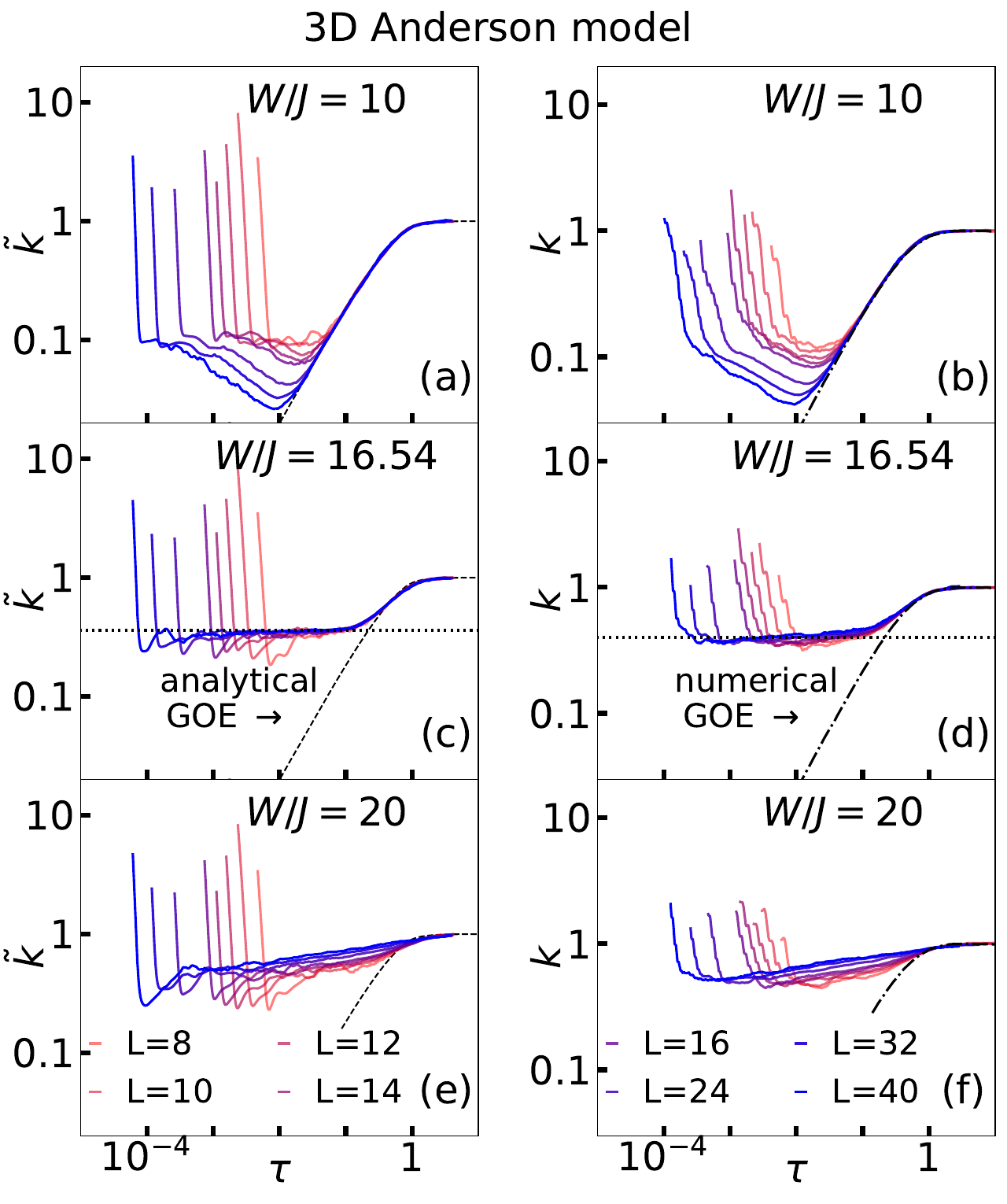}
\caption{
The SFF in the 3D Anderson model on the cubic lattice at different linear system sizes $L$.
[(a),(c),(e)] The SFF $\tilde{k}(\tau)$ after spectral unfolding and filtering, see Eq.~(\ref{def_Kt}).
[(b),(d),(f)] The SFF $k(\tau)$ without unfolding and filtering, see Eq.~(\ref{def_ktau}).
(a) and (b) $W/J=10$, (c) and (d) the critical point $W/J=16.54$, (e) and (f) $W/J=20$.
The horizontal dotted lines in (c) and (d) denote the plateau values of $\tilde{k}(\tau)$ and $k(\tau)$. 
The dashed lines in (a), (c), and (e) denote the analytical expression for the GOE ramp from Eq.~(\ref{eq:K_goe_ramp}), while the dashed-dotted lines in (b), (d), and (f) denote the numerical result for the ramp in the GOE Hamiltonians, cf.~Fig.$~$\ref{fig_3}.
}\label{fig_20}
\end{figure}

\begin{figure}[!t]
\centering
\includegraphics[width=0.98\columnwidth]{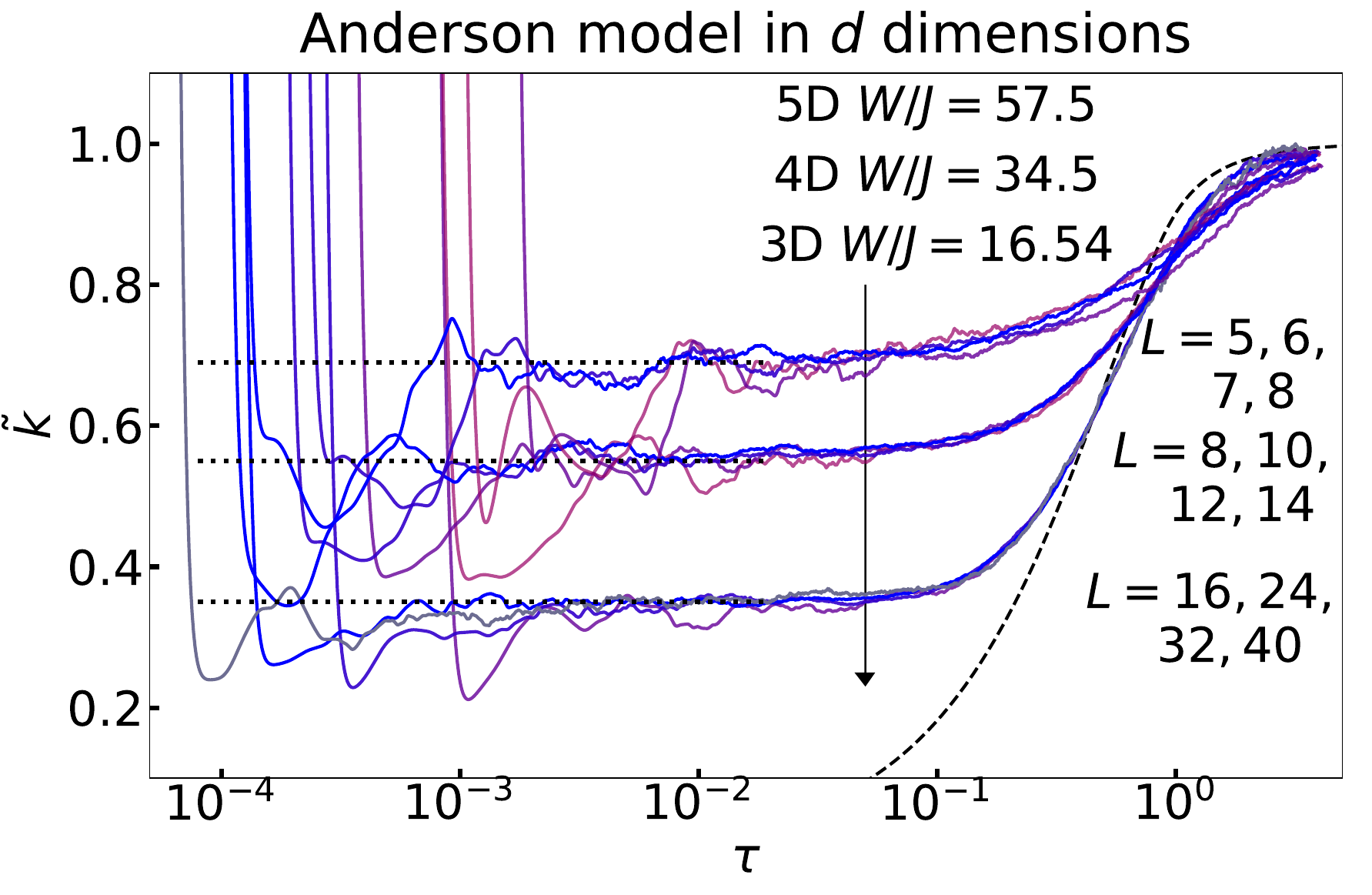}
\caption{
The SFF $\tilde{k}(\tau)$ after spectral unfolding and filtering, see Eq.~(\ref{def_Kt}), at the critical points of the Anderson models on 3D, 4D and 5D hypercubic lattices, shown at different system sizes $L$.
The dotted lines denote the plateau values of $\tilde{k}(\tau)$.
The dashed line denotes the analytical expression for the GOE ramp from Eq.~(\ref{eq:K_goe_ramp}).
}\label{fig_21}
\end{figure}

Figure$~$\ref{fig_20} compares the unfolded and filtered $\tilde{k}(\tau)$ in the 3D Anderson model, studied in~\cite{suntajs_prosen_21}, with $k(\tau)$ studied in the main text.
Results are qualitatively very similar.
In the chaotic regime, see Figs.~\ref{fig_20}(a) and~\ref{fig_20}(b), we observe that in the late-time dynamics after the Thouless time, $\tilde{k}(\tau)$ follows the analytical prediction of the GOE from Eq.~(\ref{eq:K_goe_ramp}), while $k(\tau)$ better agrees with the numerical results for the GOE Hamiltonians.
At criticality, see Figs.~\ref{fig_20}(c) and~\ref{fig_20}(d), $\tilde{k}(\tau)$ exhibits a clearer plateau in the mid-time dynamics, however, it also exhibits a dip below the plateau at short times.
We also note that the plateau value of $\tilde{k}(\tau)$ emerges at slightly higher value when compared to the plateau value of $k(\tau)$.
On the localized side, see Figs.~\ref{fig_20}(e) and~\ref{fig_20}(f), the results are again similar, apart from the dips at short times that emerge in $\tilde{k}(\tau)$.

Figure~\ref{fig_21} extends the analysis of $\tilde{k}(\tau)$ at criticality to the 4D and 5D Anderson models.
This result should be compared to Fig.~\ref{fig_11}(a), in which we show $k(\tau)$.
The tendency for the emergence of a scale-invariant plateau in the mid-time dynamics is clearer in $\tilde{k}(\tau)$ in Fig.~\ref{fig_21}.
However, $\tilde{k}(\tau)$ also exhibits some deviations when compared to the analytical prediction for the GOE ramp at late times.

\begin{figure}[!t]
\centering
\includegraphics[width=0.98\columnwidth]{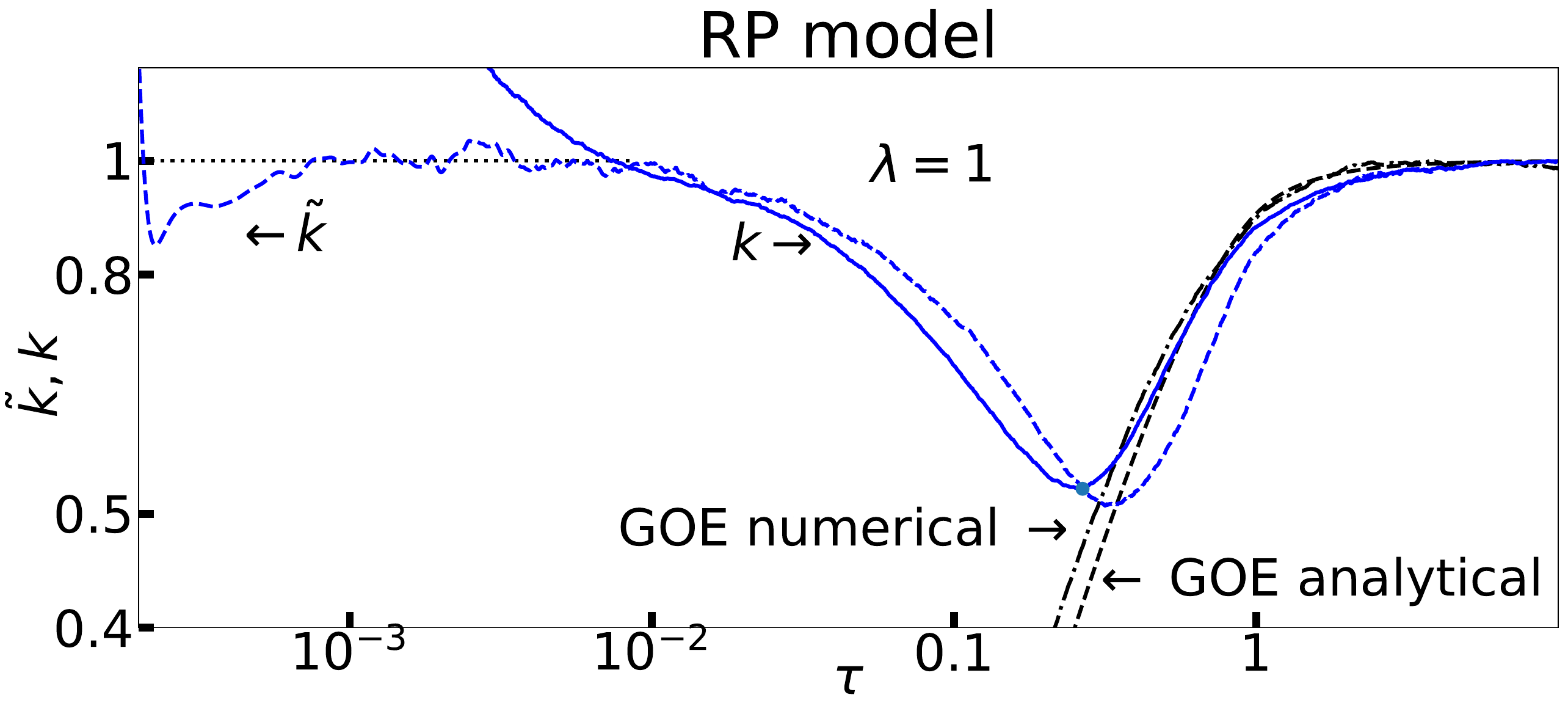}
\caption{
The SFFs $k(\tau)$ from Eq.~(\ref{def_ktau}) (solid blue line) and $\tilde{k}(\tau)$ from Eq.~(\ref{def_Kt}) (dashed blue line) in the RP model at $c=2$, $\lambda=1$ and $L=D=20000$.
The dotted lines denote the plateau values of $\tilde{k}(\tau)$ and $k(\tau)$, which are both equal to one.
The dashed black line denotes the analytical expression for the GOE ramp from Eq.~(\ref{eq:K_goe_ramp}), while the dashed-dotted black line denotes the numerical result for the ramp in the GOE Hamiltonians, cf.~Fig.$~$\ref{fig_3}.
}\label{fig_22}
\end{figure}

We also study the impact of unfolding and filtering on the SFF in the RP model.
Figure~\ref{fig_19}(a) shows $k(\tau)$ at criticality, which exhibits a tendency toward forming a plateau in the mid-time dynamics at $k_\chi = 1$.
Figure~\ref{fig_22} shows that the emergence of a plateau can be demonstrated more convincingly in $\tilde{k}(\tau)$.
On the other hand, the agreement between $\tilde{k}(\tau)$ and the analytical prediction for the GOE ramp at late times $\tau\approx 1$ is again less accurate.

\section{Details of averaging and the extraction of Thouless time} \label{app:2}

For the results shown in all figures of the paper, i.e., for all models and all system sizes, we average over minimally $100$ Hamiltonian realizations.
For smaller system sizes the number of Hamiltonian realizations is increased to $500$.
For the SFF and other survival probabilities, we additionally use the running averages in $\tau$ to reduce time fluctuations.

In several cases we extracted the Thouless time from our numerical results.
At the Thouless time $\tau_{Th}$, the SFF $k^{}(\tau=\tau_{Th})$ approaches the numerical results for the ramp $k_{\rm GOE}(\tau_{Th})$ in the GOE Hamiltonians.
We follow Refs.~\cite{suntajs_bonca_20a, suntajs_prosen_21} to establish the criterion when do the two curves approach each other.
In particular, we require
$\ln{ [k^{}(\tau_{Th})/k_{\rm GOE}(\tau_{Th})]}<\epsilon$, where the threshold $\epsilon$ is set to $\epsilon=0.05$.

\bibliographystyle{biblev1}
\bibliography{references,references1,references2}

\end{document}